\RequirePackage{fix-cm}

\documentclass{svjour3}                     
\smartqed  

\usepackage{xcolor}
\usepackage{subfig}
\usepackage{mathtools}
\usepackage{algorithm}
\usepackage{algorithmic}
\usepackage{amsfonts}
\usepackage{amssymb}
\usepackage[numbers,sort&compress]{natbib}
\usepackage{float}
\usepackage{pifont}
\usepackage{array,multirow}

\usepackage{graphicx}
\newcommand{\eqnref}[1]{(\ref{#1})}

\newcommand{\ignore}[1]{}
\newcommand{\at}[2][]{#1|_{#2}}

\DeclareMathOperator*{\argmin}{arg\,min}

\DeclarePairedDelimiterX{\inner}[2]{\langle}{\rangle}{#1, #2}

\def\mc#1{\mathcal{#1}}
\def\mb#1{\mathbb{#1}}

\def\ul#1{\underline{#1}}

\newcommand{\xmark}{\ding{55}}
\newcommand{\cmark}{\ding{51}}

\newcommand{\norm}[1]{\left\lVert#1\right\rVert}

\begin{document}

\title{Reduced-order control using low-rank Dynamic Mode Decomposition
}

\titlerunning{Reduced-order control using low-rank DMD}        

\author{Palash Sashittal         \and
        Daniel J. Bodony 
}


\institute{Palash Sashittal\at
              Department of Aerospace Engineering, \\
              University of Illinois, Urbana-Champaign\\
              \email{sashitt2@illinois.edu}           
           \and
           Daniel J. Bodony \at
              Department of Aerospace Engineering, \\
              University of Illinois, Urbana-Champaign
}

\date{Received: date / Accepted: date}

\maketitle

\begin{abstract}
In this work we perform full-state LQR feedback control of fluid flows using non-intrusive data-driven reduced-order models. We propose a model reduction method called low-rank Dynamic Mode Decomposition (lrDMD) that solves for a rank-constrained linear representation of the dynamical system. lrDMD is shown to have lower data reconstruction error compared to standard Optimal Mode Decomposition (OMD) and Dynamic Mode Decomposition (DMD), but with an increased computational cost arising from solving a non-convex matrix optimization problem. We demonstrate model order reduction on the complex linearized Ginzburg-Landau equation in the globally unstable regime and on the unsteady flow over a flat plate at a high angle of attack. In both cases, low-dimensional full-state feedback controller is constructed using reduced-order models constructed using DMD, OMD and lrDMD. It is shown that lrDMD stabilizes the Ginzburg-Landau system with a lower order controller and is able to suppress vortex shedding from an inclined flat plate at a cost lower than either DMD or OMD. It is further shown that lrDMD yields an improved estimate of the adjoint system, for a given rank, relative to DMD and OMD.
\end{abstract}

\section{Introduction}

The dynamics of many flows of engineering interest are high dimensional. Solving optimal control and design problems for these high dimensional systems is challenging, even on very large computers~\cite{kim2014adjoint}. This creates a need for reduced-order models that can faithfully approximate the dynamics of the system. Model reduction and reduced-order control therefore has been an active area of research in fluid mechanics. The reader is referred to~\cite{rowley2017model} for a thorough review of model reduction techniques for flow analysis and control.
\ignore{
\begin{itemize}
    \item data driven model reduction: we have model based methods in which a reduced basis is extracted from data and governing equations ar eprojected, review is here
    \item however this has proelbmes when eqns not known etc etc. we use system identificaiton methods are used for model order reduction
    \item reduced order control using sys identification
    \item DMD has also been used but they did it this way and I do it that way
\end{itemize}
}

Data-driven model reduction techniques use high-fidelity numerical or experimental data to extract simplified models that capture essential features of the flow. Model-based control methods, such as Galerkin projection~\cite{carlberg2011efficient,willcox2002balanced,rowley2005model}, extract a reduced basis from the snapshots and project the full system dynamics onto the linear subspace formed by the reduced basis. Reduced order feedback control is then constructed using the resulting reduced order model. Model-based feedback control has been used in~\cite{bagheri2009input,semeraro2011feedback,semeraro2013transition} to control flat-plate boundary layers and in~\cite{illingworth2016model} to suppress vortex shedding over a circular cylinder. Although these methods can be very effective (see~\cite{kim2007linear} for a thorough review), they require prior knowledge of the full-state governing equations and require computation of the projection coefficients, which is inconvenient for complex nonlinear systems.

To overcome these issues, non-intrusive system identification methods are used to construct reduced order models using data without prior knowledge of the governing equations. The authors in~\cite{huang2008control} and \cite{herve2012physics} used Autoregressive Modeling~\cite{akaike1969fitting} applied to flow over an airfoil and backward facing step respectively. In this method, a mathematical relationship between the input and output data is assumed with unknown parameters that are learned by minimizing the error between the data and model prediction. An alternative approach is to estimate the state-space model using `subspace identification methods' such as N4SID~\cite{van1994n4sid,van2012subspace} which have been used for control of transitional boundary layer in~\cite{inigo2014dynamic}. A thorough review of these methods is presented in~\cite{qin2006overview}. Finally, Eigensystem Realization Algorithm (ERA)~\cite{juang1985eigensystem}, which gives a balanced reduced order state-space model of the system, has been used extensively in flow control applications~\cite{brunton2014state,flinois2016feedback,ahuja2010feedback,illingworth2012feedback,belson2013feedback,illingworth2014active}.

\ignore{
\begin{itemize}
    \item \cite{bhattacharjee2018optimal}, Optimal actuator selection for airfoil separation control
    \item \cite{deem2018experimental}, experimental, open-loop DMD airfoil separation
    \item \cite{inigo2014dynamic}, feedforward control using N4SID on transitional boundary layer
    \item \cite{herve2012physics}, ARMAX applied to backward-facing step. use the motivation from this paper to make transition from using model-based methods to data-based methods
    \item \cite{huang2008control}, ARX used on flow separation control
    \item \cite{flinois2016feedback}, feedback control of flow over D shaped body using ERA, data based
    \item \cite{cabell2006experimental}, experimental Feedback Control of Flow-Induced Cavity Tones, LQR controllers
    \item \cite{illingworth2016model}, model based feedback control, using ERA for projection
    \item 
    \item \cite{kim2007linear}, review on model based methods of flow control
    \item \cite{qin2006overview}, overview of subspace identification methods 
\end{itemize}
}

Dynamic Mode Decomposition (DMD)~\cite{schmid2010dynamic,tu2013dynamic} is a data-driven model order reduction method that learns a linear approximation of the dynamical system such that the model prediction best fits the data snapshots in the $L_2$ norm. DMD and its variants~\cite{goulart2012optimal,chen2011h,hemati2014dynamic,hemati2017biasing,dawson2016characterizing}, have also been used for reduced-order control of systems with uncertain parameters~\cite{kramer2017feedback}, for suppressing flow separation~\cite{deem2018experimental,hemati2016improving}, and for optimal actuator selection for airfoil separation control~\cite{bhattacharjee2018optimal}. These methods approximate the dynamics of the system by fitting an endomorphic linear function on some low-dimensional subspace. The linear approximation of the dynamical system is learned constrained to a particular low dimensional subspace. This is very useful in model order reduction since it provides a low order basis to describe the flow. However, in this study, we find that this constraint is restrictive for the reduced order control performance of these methods. We investigate the possibility of using a linear map between different subspaces to construct reduced-order models and controllers for unsteady fluid flows. To do this, we formulate a rank-constrained matrix optimization problem and propose two methods to solve it. We call this model reduction method low-rank Dynamic Mode Decomposition (lrDMD)~\cite{sashittal2018low,heas2016low}. We apply full-state feedback control, constructed using the LQR framework, on model flow problems such as the Ginzburg-Landau equation and flow past an inclined flat plate. The performance of the controller is compared to controllers constructed using DMD and Optimal Mode Decomposition (OMD)~\cite{goulart2012optimal,wynn2013optimal} methods.

The outline of the paper is as follows. Section~\ref{sec:math_formulation} outlines the mathematical formulation of the problem and the three methods we investigate: DMD, OMD and lrDMD. Section~\ref{sec:numerical_method} details the numerical implementation of subspace projection and gradient descent algorithms used to solve the lrDMD optimization problem, including the model reduction computational performance while Section~\ref{sec:controller_design} describes the controller design. The control application results are shown in Section~\ref{sec:results} and Section~\ref{sec:conclusion} concludes the paper.
\section{Mathematical Formulation}\label{sec:math_formulation}

Consider a dynamical system with the state vector $x\in \mb{R}^m$ such that
\begin{align}
    x_{k+1} = f(x_k)
    \label{eq:dyn_sys}
\end{align}
where the subscript denote time iteration number. We have access to a sequence of time snapshots of this state vector which we represent as a data matrices $X, Y \in \mb{R}^{m\times n}$ formed by $n$ pairs of data snapshots as follows,
\begin{align*}
    X\coloneqq (x_1|\cdots | x_n), \quad Y\coloneqq (x_2|\cdots |x_{n+1}).
\end{align*}
\ignore{such that $x'_i = f(x_i)$ $\forall \ i\in\{1,\cdots,n\}$. }The matrix $Y$ is called the time-shifted data matrix associated with $X$. Our aim is to build a reduced-order model of the dynamical system~\eqnref{eq:dyn_sys} that captures the essential features of the function $f(\cdot)$. To this end, we construct a linear approximation of $f(\cdot)$ from the given data matrices. A natural choice of optimization problem to solve for this purpose would be
\begin{align}
    \Hat{A} = \argmin_{A\in \mb{R}^{m\times m}} \norm{Y - A X}_F^2,
    \label{eq:full_opt}
\end{align}
where $\norm{\cdot}_F$ is the Frobenius norm, $\Hat{A}\in \mb{R}^{m\times m}$ is the inferred state transition matrix. However, for fluid systems $m$ is generally large, frequently in the range of $10^6 - 10^9$, which makes this optimization problem computationally intractable. To remedy this,\ignore{ some constrained version of this problem is adopted to find a low-rank representation of the dynamical system} low rank versions of this problem are adopted. We will review two popular methods: Dynamic Mode Decomposition (DMD)~\cite{schmid2010dynamic,tu2013dynamic} and Optimal Mode Decomposition (OMD)~\cite{goulart2012optimal,wynn2013optimal}, and introduce a generalized version referred to as low-rank Dynamic Mode Decomposition (lrDMD)~\cite{sashittal2018low,heas2016low}.
\subsection{Dynamic Mode Decomposition}\label{sec:DMD}

DMD~\cite{schmid2010dynamic,tu2013dynamic} provides the optimal linear representation of the dynamical system in the space spanned by the Proper Orthogonal Decomposition (POD) modes of the data matrix $X$. POD modes are the directions that maximally capture the energy or variance in a given data matrix. For a deterministic system, and taking the $L_2$ norm of the state vector as the energy measure, the POD modes are given by the left singular vectors of the data matrix $X$ ranked by the value of the corresponding singular values. For a $r$-rank representation of the dynamics we consider the space spanned by the $r$ leading POD modes of the data matrix. Let the rank $r$ reduced singular value decomposition of $X$ be
\begin{align*}
    X \approx U\Sigma V^T,
\end{align*}
where $U\in \mb{R}^{m\times r}$, $\Sigma\in\mb{R}^{r\times r}$ and $V\in\mb{R}^{n\times r}$. Then the DMD solution $\tilde{A}\in\mb{R}^{r\times r}$ is given by
\begin{align*}
    \tilde{A} = U^TYV\Sigma^{-1}.
\end{align*}
The rank $r$ linear approximation of the state transition matrix is given by
\begin{align*}
    \hat{A} = U\tilde{A}U^T.
\end{align*}
\subsection{Optimal Mode Decomposition}\label{sec:OMD}

In OMD~\cite{goulart2012optimal,wynn2013optimal}, the following optimization problem is solved to give an approximate solution of \eqref{eq:full_opt},
\begin{align}\label{eq:omd_opt}
    \min_{L,M} &\norm{Y - LML^TX}_F^2 \\
    \text{s.t.} \; & L^TL = I \\ 
                   & M\in\mb{R}^{r\times r}, L\in\mb{R}^{m\times r}.
\end{align}
where $r$ is the rank of the linear approximation of the state transition matrix $\hat{A} = LML^T$. This particular form of low rank approximation has useful implications in reduced-order modeling. The matrix $L$ provides the basis of a low dimensional subspace to approximate the trajectories of the dynamical system while $M$ provides the dynamical evolution of the state vector on this subspace. The solution of $M$ for a fixed $L$ is given by
\begin{align*}
    M^*(L) = L^TYX^TL(L^TXX^TL)^{-1}.
\end{align*}
The OMD solution coincides with the DMD approximation when $L$  is chosen as the $r$ leading left singular vectors of the data matrix $X$. In this way OMD is a generalization of the DMD formulation.
\subsection{Low-Rank Dynamic Mode Decomposition}\label{sec:lrDMD}

We consider a rank-constrained solution of the optimization problem in~\eqref{eq:full_opt}. Recently the authors of~\cite{heas2016low} showed that there exists a closed form solution for this optimization problem. We solve the following equivalent optimization problem,
\begin{align}
    \min_{L,D,R} \norm{Y - LDR^TX}_F^2,
    \label{eq:prob}
\end{align}
where $L,R\in \mb{R}^{m\times r}$ and $L^TL=R^TR=I_r$ ($r\times r$ identity matrix) and $\hat{A} = LDR^T$ is the $r$-ranked matrix approximating the dynamics of the underlying system. Observe that \eqref{eq:prob} optimizes over all possible left and right subspaces of the state transition matrix in contrast to standard DMD and OMD. A key step to solving \eqref{eq:prob} is to find the optimal solution for $D$ as a function of $L$ and $R$. For a fixed $L$ and $R$ there exists a closed form optimal solution for $D$, namely
\begin{align}
    D^*(L,R) = (L^TYX^TR)(R^TXX^TR)^{-1}.
    \label{eq:D}
\end{align}
Using \eqref{eq:D} in the objective function and simplifying further the optimization problem can be reduced to
\begin{align}
    \min_{L,R}\; \left(- \norm{L^TYQ_R}_F^2\right)
    \label{eq:final_prob}
\end{align}
where $Q_{R} = X^TR(R^TXX^TR)^{-1}R^TX$. We will denote the objective function~\eqref{eq:final_prob} by $G(L,R)$. An important observation is that $G(L,R)$ is only a function of the spaces spanned by the columns of $L$ and $R$. To see this consider an orthogonal matrix $P\in \mb{R}^{r\times r}$ such that $P^TP = I_r$. For this objective function, $G(LP,R) = G(L,R)$ and $G(L,RP) = G(L,R)$. This implies that the objective function does not change values as long as the space spanned by the columns of $L$ and $R$ do not change. Hence the optimization can be equivalently performed over the set of $r$-dimensional subspaces in $\mb{R}^m$ instead of the set of orthogonal matrices in $\mb{R}^{m\times r}$.

The set of $r$-dimensional subspaces in $\mb{R}^m$ is known as the Grassmanian manifold $\mc{G}_{r,m}$~\cite{absil2009optimization,edelman1998geometry}. In matrix representation an element of $\mc{G}_{r,m}$ is specified by an orthogonal basis of the $r$-dimensional subspace. The manifold on which we perform the optimization~\eqref{eq:final_prob} is a product manifold of two Grassmanian manifolds
$$\mc{M} \coloneqq \mc{G}_{r,m}\times \mc{G}_{r,m}.$$ 
An element of $\mc{M}$, in matrix representation, will be characterized by $M = (L,R)$.

\begin{remark}
The $L$ and $R$ basis solutions of the optimization problem~\eqref{eq:final_prob} need not be the best basis for a Petrov-Galerkin reduced-order model. These bases only provide the subspaces for the best low-rank linear description of the dynamical system in the least-squares sense. As will be made clear in Section~\ref{sec:controller_design}, for the reduced-order model we use the input basis $R$ for both the trial subspace and the test subspace.   
\end{remark}
\section{Numerical Methods for lrDMD optimization problem}\label{sec:numerical_method}

In this section we present the numerical methods employed to solve the lrDMD optimization problem and compare the computational performance of the methods to OMD and DMD. Recall that the objective function and the optimization problem are,
\begin{align}\label{eq:opt_prob}               
    \min_{L,R}\; & G(L,R) \coloneqq \left(- \norm{L^TYQ_R}_F^2\right) \\
    \text{s.t.} \; & (L,R) \in \mc{M} \coloneqq \mc{G}_{r,m}\times\mc{G}_{r,m} \nonumber\\ 
    \text{with} \; & Q_{R} = X^TR(R^TXX^TR)^{-1}R^TX.\nonumber
\end{align}
Two methods to solve this optimization problem are detailed in Appendix~\ref{app:methods}. The first method in Appendix~\ref{app:sub_proj} is a computationally efficient subspace projection method that provides an approximate solution to the problem. The second method in Appendix~\ref{app:gradD} is a gradient-based method that is guaranteed to converge to a local minimum.

\begin{remark}
There is no guarantee that the gradient based method will converge to the global minimizer of the given optimization problem~\eqref{eq:opt_prob}. The gradient based method can therefore be thought as a post-processor for the initial guess provided by either DMD or the subspace projection method described in Algorithm~\ref{alg:subspace}.
\end{remark}

\subsection{Computational performance}\label{sec:comp_perf}

The advantage to using the form of the optimization problem in Eq.~\eqref{eq:prob} instead of Eq.~\eqref{eq:full_opt} is that we only need to solve for $\mc{O}(m)$ variables for the fully parameterized state transition matrix. For large $m$, even this can make the problem computationally intractable. An important observation that can help is that we only need to consider the solution of $R$ such that $\text{Im}(R) \subseteq \text{Im}(X)$. Consider $M$ to be the full ranked data matrix that contains all the data snapshots that form the data matrices $X$ and $Y$ and let there be $p$ such snapshots. Consider the singular value decomposition of the data matrix $M \in \mb{R}^{m\times p}$
\begin{align*}
    M = U\Sigma V^T.
\end{align*}

We assume that the basis of both $L$ and $R$ is contained within $U$ so that
\begin{align*}
    L &= U\ul{L} \\
    R &= U\ul{R}
\end{align*}
where $\ul{L},\ul{R} \in \mc{G}_{r,p}$. The optimization problem reduces to
\begin{align*}
    \min_{\ul{L},\ul{R}}\; & G(\ul{L},\ul{R}) \coloneqq (- \norm{\ul{L}^T\ul{Y}Q_{\ul{R}}}_F^2) \\
    \text{s.t.} \; & (\ul{L},\ul{R}) \in \mc{M} \coloneqq \mc{G}_{r,p}\times\mc{G}_{r,p} \\
               & Q_{\ul{R}} = \ul{X}^T\ul{R}(\ul{R}^T\ul{X}\ul{X}^T\ul{R})^{-1}\ul{R}^T\ul{X}
\end{align*}
where $\ul{Y} = U^TY$ and $\ul{X} = U^TX$. Since $p \ll m$, this saves significant computational time. In this study, we use this technique for lrDMD and OMD where gradient based methods are used to solve the optimization problem. Therefore, in this study, OMD and the gradient based implementation of lrDMD solve the optimization problem with the reduced data matrices $\ul{Y}$ and $\ul{X}$.

\begin{table}
    \centering
        \begin{tabular}{|c|c|c|c|c|c|c|}
            \hline
             \multirow{2}{4em}{Method} & \multicolumn{3}{|c|}{$n = 50$} & \multicolumn{3}{|c|}{$n = 200$} \\
             \cline{2-7}
                 &  $r = 10$ &  $r = 20$ & $r = 30$ &  $r = 10$ &  $r = 20$ & $r = 30$ \\
             \hline
             DMD & $0.1443$ & $0.1317$ & $0.1096$ & $0.6965$ & $0.6691$ & $0.6464$ \\
             \hline
             OMD & $3.2687$ & $5.5517$ & $10.5841$ & $1.6759$ & $17.4687$ & $26.5706$ \\
             \hline
           lrDMD (subProj) & $0.7731$ & $0.4127$ & $1.4995$ & $2.3426$ & $2.8213$ & $1.4139$\\
             \hline
           lrDMD (GradD)  & $19.9936$ & $2.8781$ & $0.8261$ & $58.0509$ & $17.1708$ & $26.0866$\\
           \hline
        \end{tabular}
    \caption{Time taken in seconds by DMD, OMD and lrDMD for generating reduced-order models of rank $r$ with data matrices composed of $n$ snapshots. lrDMD (subProj) refers to subspace projection method implemented using Algorithm~\ref{alg:subspace} and lrDMD (GradD) refers to gradient based method described in detail in~\cite{sashittal2018low}}
    \label{tab:time_taken}
\end{table}

Table~\ref{tab:time_taken} compares the computational performance of DMD, OMD and the two algorithms to solve the lrDMD problem. It shows the time taken in seconds by the three methods in computing a reduced-order model for the given data matrices. The data snapshots have dimension $m = 62001$ and describe the vorticity field for a flow past flat plate described in Section~\ref{sec:ibpm} with consecutive snapshots that are $100$ timesteps apart. The data matrices are generated with $n = \{50, 200\}$ snapshots and reduced-order models of rank $r = \{10, 20, 30\}$ are computed. lrDMD (subProj) refers to the solution of lrDMD using the subspace projection method described in Algorithm~\ref{alg:subspace} whereas lrDMD (GradD) refers to the gradient based method~\cite{sashittal2018low}. The gradient based methods used in OMD and lrDMD are implemented using ManOpt~\cite{boumal2014manopt} package on MATLAB using the trust-region~\cite{absil2007trust} algorithm. The initial condition for OMD is given by the DMD solution whereas the initial condition for the lrDMD is the solution from Algorithm~\ref{alg:subspace}. All computations are performed on a standard desktop PC with a 2.2 GHz quad-core Intel i7 processor and 16GB RAM running on Mac OS X 10.11.

Table~\ref{tab:time_taken} shows that time taken by DMD is always lower than OMD or either method of solving the lrDMD problem, with the subspace projection method incurring computational times $2-3\times$ larger than DMD but is faster than OMD for all cases. An interesting observation is that for certain cases like $(n,r) = (50,30)$, the initial condition provided to `lrDMD (GradD)' by the subspace projection method is very close to the optimal and no iterations of the gradient based method are required for convergence. In almost all cases lrDMD with gradient descent takes around the same time as OMD except a few cases when OMD takes less time than lrDMD.

\begin{table} 
    \centering
        \begin{tabular}{|c|c|c|c|c|c|c|}
            \hline
             \multirow{2}{4em}{Method} & \multicolumn{3}{|c|}{$n = 50$} & \multicolumn{3}{|c|}{$n = 200$} \\
             \cline{2-7}
                 &  $r = 10$ &  $r = 20$ & $r = 30$ &  $r = 10$ &  $r = 20$ & $r = 30$ \\
             \hline
             DMD & $14.7449$ & $5.7669$ & $3.3224$ & $24.0859$ & $2.2424$ & $0.3296$ \\
             \hline
             OMD & $4.2816$ & $0.0921$ & $0.0049$ & $12.1151$ & $0.6408$ & $0.0176$\\
             \hline
           lrDMD (subProj) & $1.8183$ & $0.0160$ & $0.0015$ & $6.3463$ & $0.5206$ & $0.0295$\\
             \hline
           lrDMD (GradD)  &  $1.2686$ & $0.0031$ & $0.0015$ & $5.0654$ & $0.1638$ & $0.0058$\\
           \hline
        \end{tabular}
    \caption{Error norm $\epsilon$ of reconstruction for reduced-order models of rank $r$ generated by DMD, OMD and lrDMD with data matrices composed of $n$ snapshots. lrDMD (subProj) refers to subspace projection method implemented using Algorithm~\ref{alg:subspace} and lrDMD (GradD) refers to gradient based method described in detail in~\cite{sashittal2018low}.}
    \label{tab:my_norm}
\end{table}

\ignore{
\begin{table} 
    \centering
        \begin{tabular}{|c|c|c|c|c|c|c|c|c|c|}
            \hline
             \multirow{2}{4em}{Method} & \multicolumn{3}{|c|}{$n = 50$} & \multicolumn{3}{|c|}{$n = 100$} & \multicolumn{3}{|c|}{$n = 200$} \\
             \cline{2-10}
                 &  $r = 10$ &  $r = 20$ & $r = 30$ &  $r = 10$ &  $r = 20$ & $r = 30$ &  $r = 10$ &  $r = 20$ & $r = 30$ \\
             \hline
             DMD & $14.7449$ & $5.7669$ & $3.3224$ & $17.6765$ & $2.1319$ & $0.4041$ & $24.0859$ & $2.2424$ & $0.3296$ \\
             \hline
             OMD & $4.2816$ & $0.0921$ & $0.0049$ & $10.6180$ & $0.1741$ & $0.0051$ & $12.1151$ & $0.6408$ & $0.0176$\\
             \hline
           lrDMD (subProj) & $1.8183$ & $0.0160$ & $0.0015$ & $7.2776$ & $0.1822$ & $0.0041$ & $6.3463$ & $0.5206$ & $0.0295$\\
             \hline
           lrDMD (GradD)  &  $1.2686$ & $0.0031$ & $0.0015$ & $3.2409$ & $0.0194$ & $0.0041$ & $5.0654$ & $0.1638$ & $0.0058$\\
           \hline
        \end{tabular}
    \caption{Error norm $\epsilon = \norm{Y - \hat{A}X}_F$ of reconstruction for reduced-order models of rank $r$ generated by DMD, OMD and lrDMD with data matrices composed of $n$ snapshots. lrDMD (subProj) refers to subspace projection method implemented using algorithm~\ref{alg:subspace} and lrDMD (GradD) refers to gradient based method described in detail in~\cite{sashittal2018low}.}
    \label{tab:my_norm}
\end{table}
}

Table~\ref{tab:my_norm} shows the error norm defined by $\epsilon = \norm{Y - \hat{A}X}_F$ for all the cases considered in Table~\ref{tab:time_taken}. It can be seen that lrDMD with gradient descent and subspace projection shows error norm much less than OMD or DMD for all cases considered. Subspace projection in particular, shows lesser error than OMD in almost all cases while taking lesser time to compute the reduced-order model. For data snapshots that are spaced far apart, lrDMD with gradient descent is recommended for accurate low rank reconstruction of the system dynamics if computation time is not a factor.
\section{Controller Design}\label{sec:controller_design}

Once the reduced order model is learned, the next step is to construct the feedback control law using the reduced order model. We use a Linear Quadratic Regulator~\cite{kwakernaak1972linear} (LQR) to construct a stabilizing feedback control independent of the initial condition. Consider a linear system with the state variable $q\in\mb{R}^{m}$ and control input $u\in\mb{R}^p$. The solution of the LQR problem provides a feedback gain $K\in\mb{R}^{p\times m}$ such that the control input $u_k = -Kq_k$ minimizes
\begin{align*}
    J(q,u) = &\sum_{k=0}^{\infty} (q_k^TQq_k + u_k^TSu_k), \\
    \text{such that}\quad & q_{k+1} = Aq_{k} + B u_{k},
\end{align*}
where the subscripts denote the temporal iteration number, $A\in\mb{R}^{m\times m}$ is the state transition matrix of the linear system and $B\in\mb{R}^{m\times p}$ describes the actuator. Under the assumption that $(A,B)$ is stabilizable, the feedback gain matrix $K$ can be computed by solving the discrete  algebraic Riccati equation
\begin{align}
    A^TPA - P - (A^TPB)(S + B^TPB)^{-1}(B^TPA) + Q = 0,
    \label{eq:dare}
\end{align}
for $P$ and using $K = (S + B^TPB)^{-1}B^TPA$. Although most fluid flow systems are nonlinear, the LQR is a powerful tool used for stabilization about an unstable base flow when a feedback form of controller can be employed.

Direct methods of solving the Riccati equation~\eqref{eq:dare} have time complexity of $O(m^{6})$. As noted in the introduction, fluid flows are governed by partial differential equations which result in high dimensional systems upon discretizations, making solving Eq.~\eqref{eq:dare} computationally infeasible for practical fluid flow applications. To remedy this, we employ a Galerkin projection based method~\cite{simoncini2016analysis,alla2017order} that yields a low-rank approximation of the solution to Eq.~\eqref{eq:dare}. Let $E$ be the residual of the discrete Riccati equation as follows,
\begin{align*}
    E(P) = A^TPA - P - (A^TPB)(S + B^TPB)^{-1}(B^TPA) + Q.
\end{align*}
The projection method works by restricting the solution in some subspace and imposing an orthogonality of the residual to that subspace. Consider a $k$-dimensional subspace spanned by the basis $W\in\mb{R}^{m\times k}$. The solution of Eq.~\eqref{eq:dare} is approximated as,
\begin{align*}
    \bar{P} = WPW^T,
\end{align*}
and the orthogonality condition on the residual is $W^TE(\bar{P})W = 0$. For all the reduced-order models in this study, we choose $W$ to be some basis of the row space of the low-rank approximation of $A$. This is given by the POD basis in case of DMD, the optimal solution for $L$ in case of OMD and the optimal solution of $R$ in case of lrDMD. This choice permits a natural comparison between the three reduction methods.
\section{Results}\label{sec:results}

In this section we compare the performance of the reduced-order LQR controllers built using DMD, OMD and  lrDMD on unsteady dynamical systems. The first system we look into is the complex linear Ginzburg-Landau equation. We use the reduced-order models to find optimal feedback control as well as the optimal actuator location for the system in the unstable regime. We also employ LQR controllers on the incompressible flow past inclined flat plate at high angle of attack. The results show that the reduced-order linear feedback controllers can be effective in regions of phase-space with strong nonlinearities and can be used to suppress nonlinear vortex shedding.
\subsection{Linearized Ginzburg Landau Equation}\label{sec:ginz_landau}

In this section we apply the control strategies described so far on the linearized complex Ginzburg-Landau (GL) equation, which is a well known model equation for fluid systems. Even though all the analysis in this paper has been on the real number field, the results can be easily extended complex numbers. The GL equation is as follows,
\begin{align*}
    &\frac{\partial q}{\partial t} + \nu\frac{\partial q}{\partial x} = 
    \mu(x)q + \gamma \frac{\partial^2 q}{\partial x^2}, \\
    \text{with}\quad &\mu(x) = \mu_0 - c_u^2 + \mu_2x^2/2,
\end{align*}
where the real part of $q(x,t)$ represents velocity or stream function perturbation amplitude. The GL equation exhibits a variety of stability behaviors observed in fluid flows for different values of the constants $\nu,\, c_u,\, \mu_0$ and $\mu_2$ in different regions of the spatial domain. For this study we have taken the value of these constants corresponding to the supercritical, globally unstable regime case in \cite{chen2011h}, shown in Table~\ref{tab:gl_values}.
\begin{table}[H]
\begin{center}
\begin{tabular}{|c|c|c|c|}
\hline
 variable& description         & value \\
\hline
   $U$    & advection velocity  & $2$\\
\hline
   $c_u$  & most unstable wavenumber & $0.2$\\
\hline
   $c_d$  & dispersion parameter & $-1.0$\\
   \hline
   $\mu_0$ & overall amplification & $0.41$\\
   \hline
   $\mu_2$ & degree of non-parallelism & $-0.01$\\
   \hline
   $\mu_t$ & transitional $\mu$ & $0.32$\\
   \hline
   $\mu_c$ & critical $\mu$ & $0.4$\\
   \hline
\end{tabular}
\end{center}
\caption{Ginzburg-Landau equation parameter values corresponding to the supercritical, globally unstable regime case in~\cite{chen2011h}.}
\label{tab:gl_values}
\end{table}
In discrete space we use the spectral formulation of the derivative operators using Hermite polynomials evaluated at the Hermite nodes~\cite{bagheri2009input}. Time discretization is performed using forward Euler time-stepping scheme. In this study we consider $n=220$ grid points with the computational domain $x\in[-85,85]$ and a time-step of $dt = 1$. For more details about the system the reader is referred to \cite{bagheri2009input}. 

We will describe this system as a discrete input-output system as,
\begin{align}
    q_{k+1} = Aq_k + Bu_k,
    \label{eq:sys_ginzlandau}
\end{align}
where $q_k \in \mb{R}^{m}$ is the state variable, $A \in \mb{R}^{m\times m}$ is the state transition matrix, $B \in \mb{R}^{m\times p}$ is the spatial support of the controller and $u_k \in \mb{R}^p$ is the control input. Our goal is to find $K \in \mb{R}^{p\times m}$ such that when $u_k = -Kq_k$, we minimize the following cost,
\begin{align}
    J = \sum_{k=1}^{\infty} (q_{k}^*Qq_{k} + u_{k}^*Su_{k}),
    \label{eq:gl_cost}
\end{align}
for given positive definite matrices $Q$ and $S$. We denote the conjugate transpose of $q$ by $q^*$. The controller and the values of $Q$ and $S$ are taken from~\cite{chen2011h}. The controller we choose is a Gaussian centered in the convectively unstable region of flow, 
\begin{align}
    B = \exp\left(-\frac{(x - x_a)^2}{2\sigma^2}\right),
    \label{eq:actuator_support}
\end{align}
where we choose $x_a = 8$ and $\sigma = 5$ which places the actuator just inside the region of amplification which is $[-8.6,8.6]$. Using $15$ snapshots of the impulse response of the flow, we generate a low-rank approximation of $A$ which is used to construct reduced-order controllers using the LQR framework and to find the optimal actuator location. 


\subsubsection{Adjoint Reconstruction}

It is known that state transition matrices arising from the discretization of the linearized governing equations for various fluid flow applications are \emph{nonnormal}~\cite{schmid2012stability}. Nonnormal systems exhibit non-orthogonal eigenmodes which can differ significantly from the adjoint modes~\cite{schmid2014analysis} that are known to play a major role in flow control and optimization~\cite{luchini2014adjoint}. Therefore, it is crucial for effective flow control applications that the reduced order models not only accurately predict the flow field but also extract adjoint information from the data. 

\begin{figure} 
\centering
\subfloat[]{\label{fig:true_eigenvalues}\includegraphics[width = 0.45\textwidth]{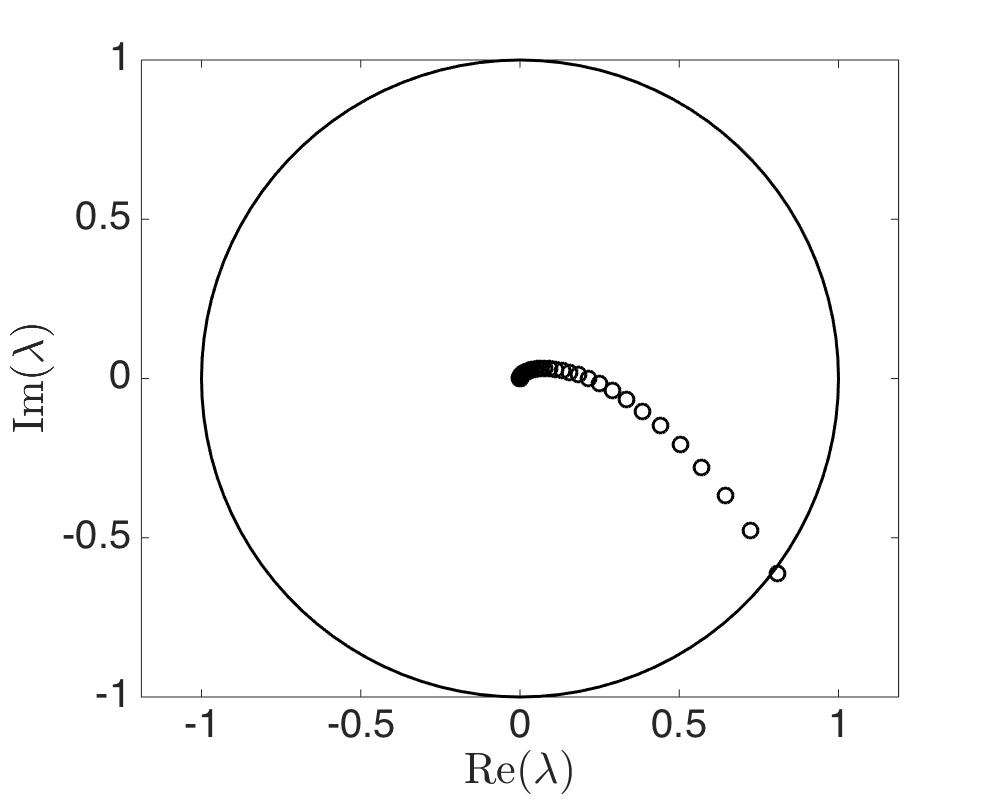}}
\subfloat[]{\label{fig:unstable_modes}\includegraphics[width = 0.45\textwidth]{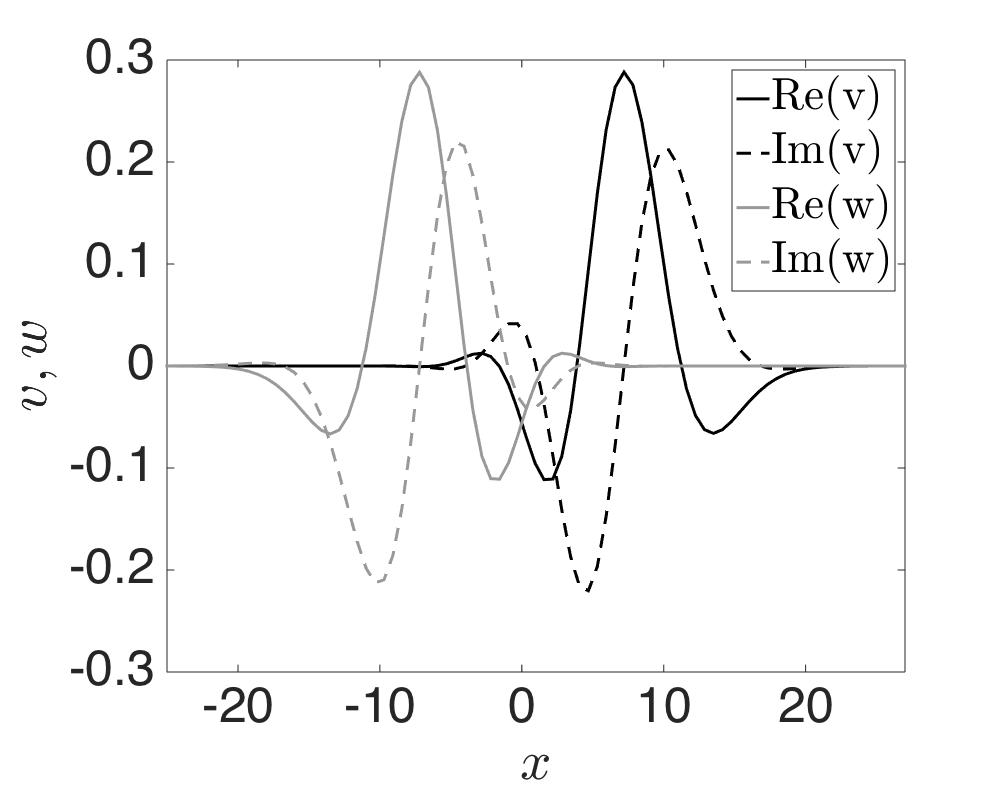}}\\
\subfloat[]{\label{fig:v-error}\includegraphics[width = 0.45\textwidth]{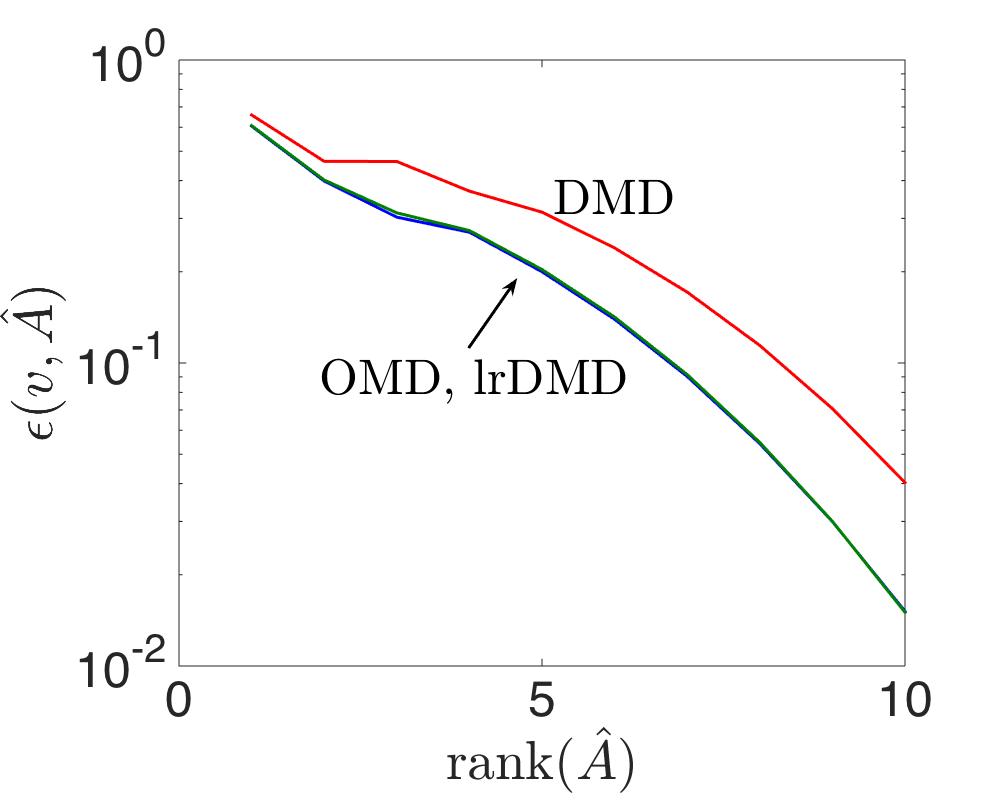}}
\subfloat[]{\label{fig:w-error}\includegraphics[width = 0.45\textwidth]{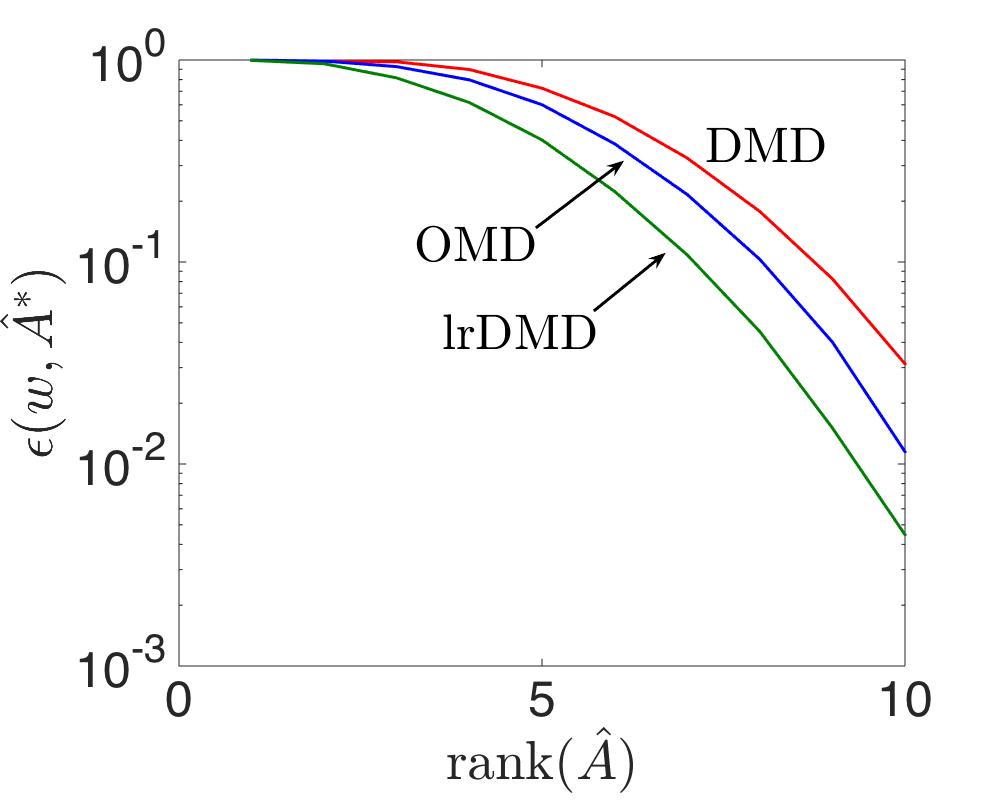}}\\
\caption{(a) Eigenvalues of the discrete GL system matrix (black circles). Unit circle is shown with a solid black line for comparison. (b) Eigenmode `$v$' and adjoint mode `$w$' corresponding to the unstable eigenvalue outside the unit circle. (c) Projection error for the unstable eigenmode for DMD (red), OMD (blue) and lrDMD (green) for different rank approximations (d) Projection error for the unstable adjoint mode for DMD (red), OMD (blue) and lrDMD (green) for different rank approximations.}
\label{fig:lrDMD_adjoint}
\end{figure}

Figure~\ref{fig:true_eigenvalues} shows the eigenvalues of the discretized state transition matrix of the GL equation in the supercritical regime. It shows one unstable eigenvalue  $0.8073 - 0.6109i$ outside the unit circle. Figure~\ref{fig:unstable_modes} shows the eigenmode $v$ and the adjoint mode $w$ corresponding to this unstable eigenvalue. For faithful reconstruction of these unstable modes, the eigenmode must lie in the column space of the reduced order model $\hat{A}$ and the adjoint mode must lie in the row space. We define the projection error
\begin{align*}
    \epsilon(q,V) = \frac{\norm{q - \mb{P}_Vq}_2}{\norm{q}_2}
\end{align*}
where $\mb{P}_V$ is the orthogonal projection matrix in the column space of $V$. We look at the projection error of the unstable eigenmode in the column space of $\hat{A}$ given by $\epsilon(v,\hat{A})$ and the projection error of the unstable adjoint mode in the row space of $\hat{A}$ given by $\epsilon(w,\hat{A}^*)$. For the DMD-based reduced order model, bases for the row space and column space are both given by the POD modes. For OMD, the solution $L$ of the optimization problem in Eq.~\eqref{eq:omd_opt} serves as the basis of both the column and the row space of the reduced order model. In the case of lrDMD, the basis of the row space is given by $R$ and the basis of the column space is given by $L$ from the solution of the optimization problem in Eq.~\eqref{eq:final_prob}.

Figure~\ref{fig:v-error} shows the projection error of the unstable eigenmode for the reduced order models of different ranks obtained from DMD, OMD and lrDMD. At all rank approximations the error incurred by OMD and lrDMD are about the same and less than the error incurred by DMD. Similarly, Figure~\ref{fig:w-error} shows the projection error of the unstable adjoint mode. In this case we see that lrDMD outperforms both OMD and DMD at all ranks. This shows that the additional degree of freedom of choosing separate input and output spaces in lrDMD allow for lower projection error in the adjoint modes of interest while also keeping the projection error in the eigenmodes as low as in OMD. 

\subsubsection{Optimal Control}

In the supercritical regime, the GL system is globally unstable, with a single eigenvalue of the state transition matrix that lies outside the unit circle, as shown in Figure~\ref{fig:true_eigenvalues}. The goal of the feedback controllers is to stabilize the system while minimizing the cost in Eq.~\eqref{eq:gl_cost}.

\begin{figure} 
\centering
\subfloat[]{\label{fig:cp_rank5}\includegraphics[width = 0.45\textwidth]{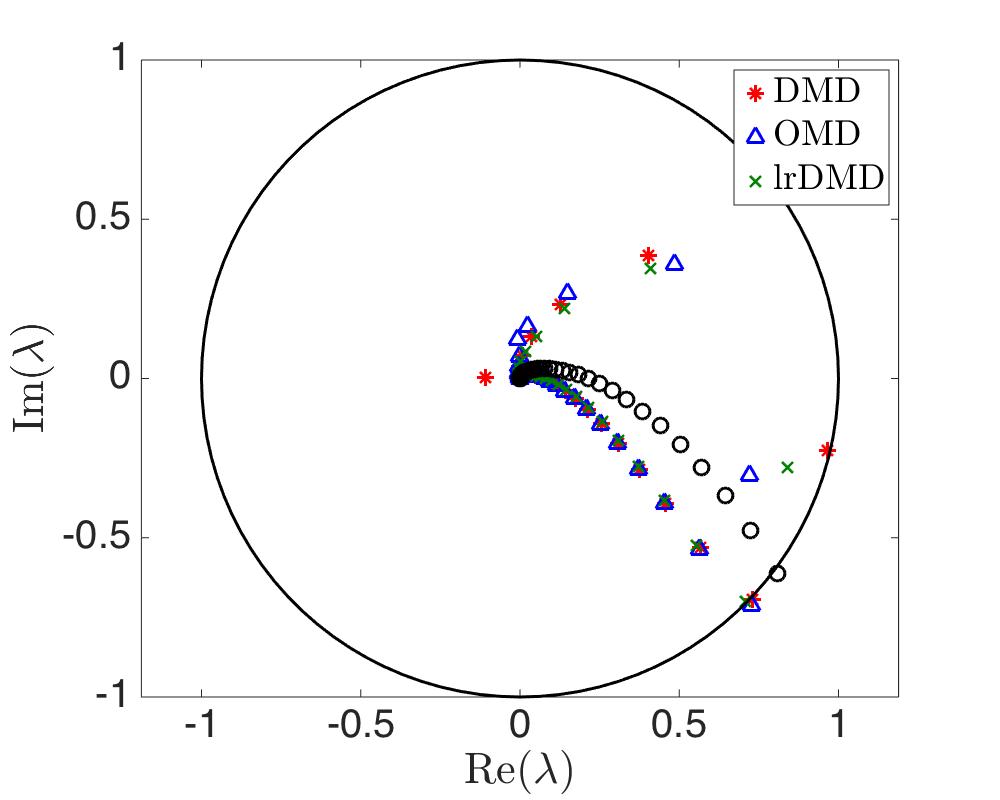}}
\subfloat[]{\label{fig:cp_rank5_zoomed}\includegraphics[width = 0.46\textwidth]{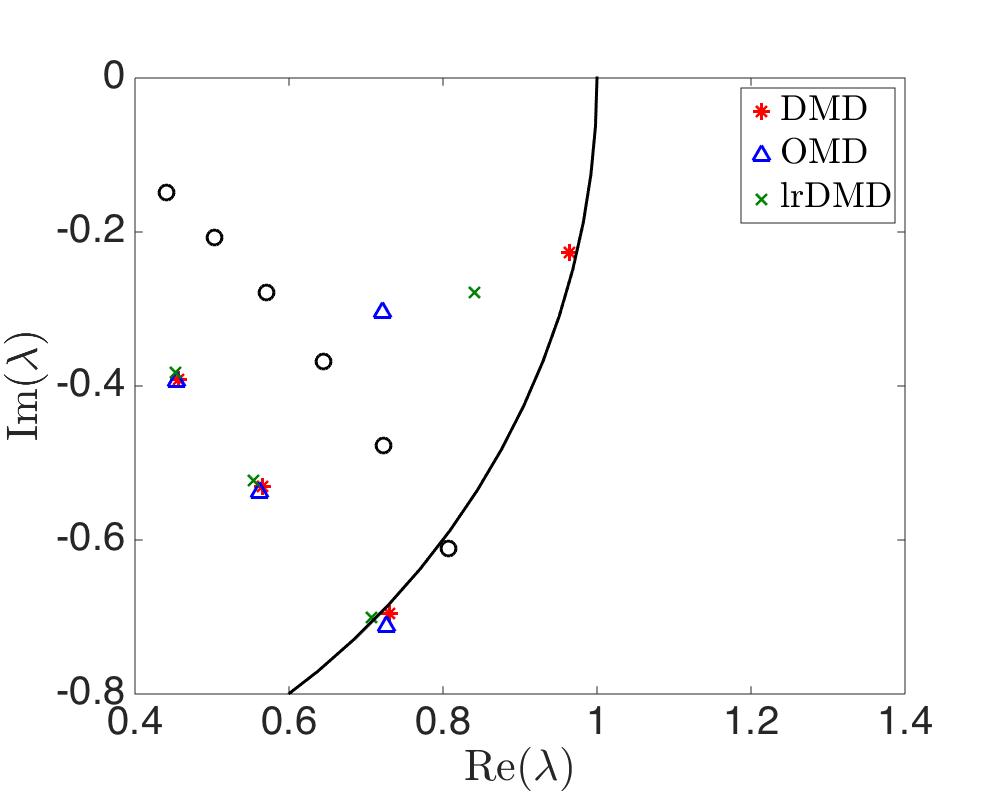}}\\
\subfloat[]{\label{fig:cp_rank9}\includegraphics[width = 0.45\textwidth]{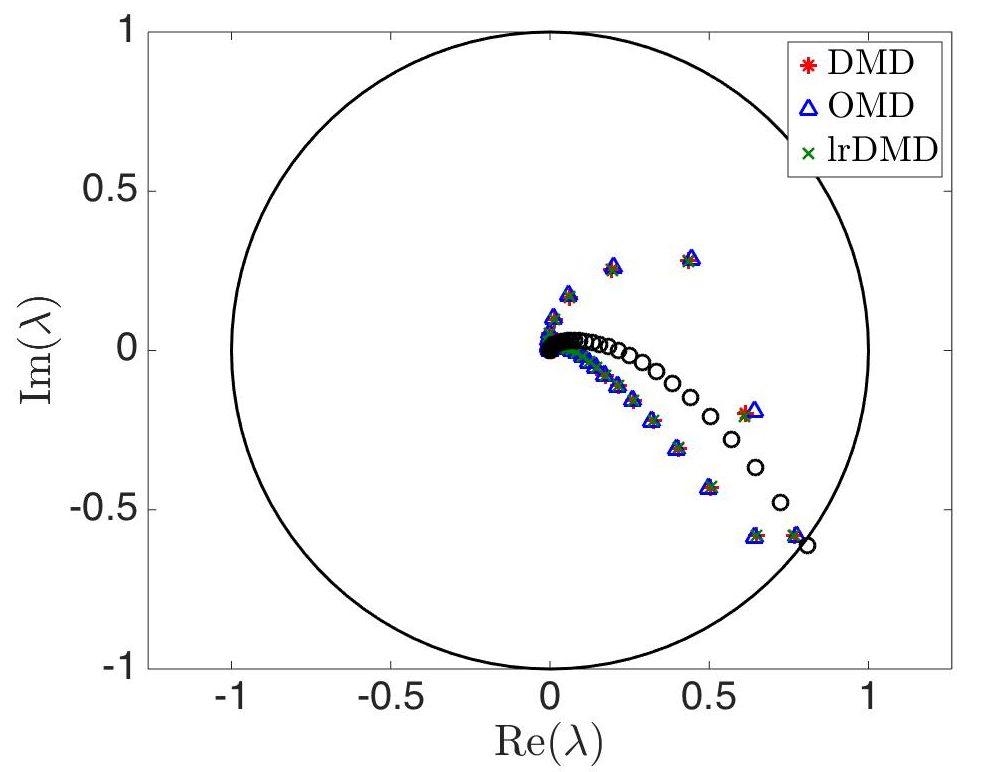}}
\subfloat[]{\label{fig:cp_true}\includegraphics[width = 0.45\textwidth]{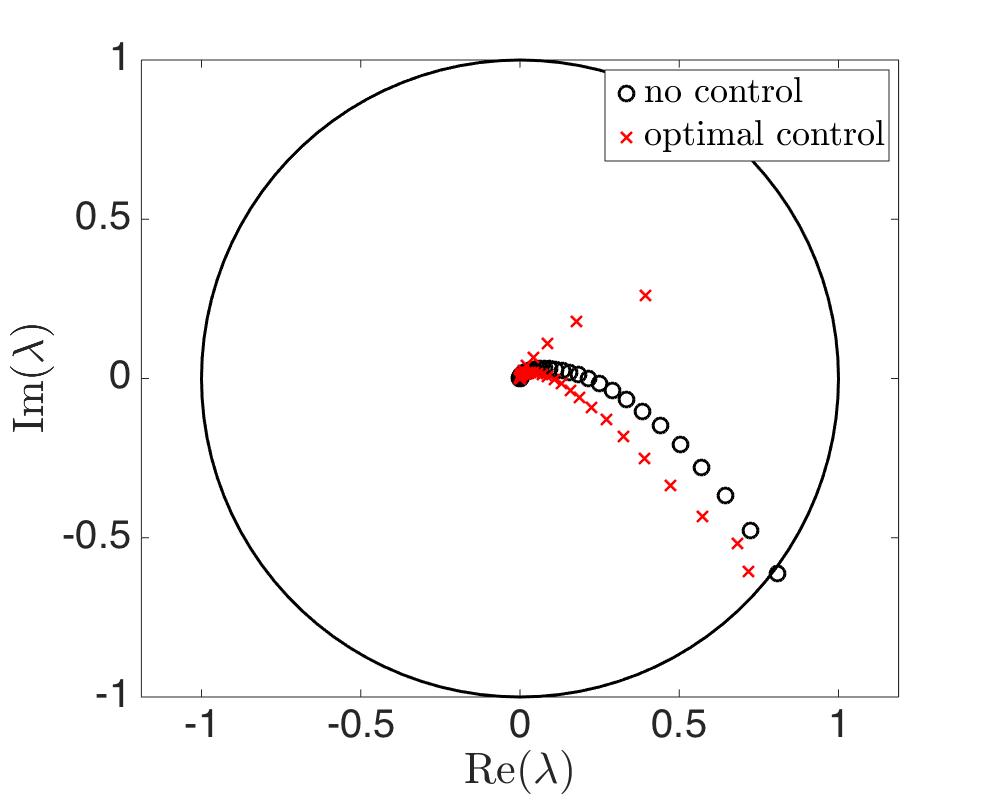}}\\
\caption{The eigenvalues of the uncontrolled ($\circ$) and controlled GL system with reduced-order controllers constructed using DMD, OMD and lrDMD with (a,b) rank 5 and (c) rank 9 approximations. (d) Shows the eigenvalues of uncontrolled and optimally controlled GL system.}
\label{fig:control_performance}
\end{figure}

Figure~\ref{fig:control_performance} shows the eigenvalues of the controlled Ginzburg-Landau system using reduced-order LQR controllers constructed using DMD, OMD and  lrDMD. Figures~\ref{fig:cp_rank5} and~\ref{fig:cp_rank5_zoomed} illustrate the rank 5 approximations whereas Fig. ~\ref{fig:cp_rank9} shows the rank 9 approximations. Figure~\ref{fig:cp_true} shows the eigenvalues of the controlled GL system with LQR optimal control constructed using the full-order system matrix. It can be clearly seen in Fig.~\ref{fig:cp_rank5_zoomed} that for the rank 5 approximation, DMD is not able to stabilize the system and has one eigenvalue outside the unit circle. OMD also fails to stabilize the system with one eigenvalue outside the unit circle. This shows that $5$ modes are not sufficient for reduced order control even with OMD which finds the optimal basis for the low order subspace. However, the freedom of choosing separate input and output subspaces enables lrDMD based reduced order controller to stabilize the system at the same rank $5$ approximation. For the rank $9$ approximation in Fig.~\ref{fig:cp_rank9}, we see that the DMD, OMD and lrDMD controlled system eigenvalues are very close to each other and also close to the eigenvalues of the optimally controlled system. Thus with sufficient rank approximation all three methods can stabilize the system but lrDMD can stabilize the system at a rank lower than OMD or DMD.
\subsubsection{Optimal actuator location}

Optimal actuator placement is a challenging problem in stability theory especially when applied to large scale fluid systems. Reduced order models make the investigation of optimal actuator location computationally feasible. We use the reduced order feedback controllers to find the optimal actuator location $x_a$ (in Eq.~\ref{eq:actuator_support}) in the computational domain for a fixed variance in the spatial support of the actuator. The authors in~\cite{chen2011h} find the optimal actuator and sensor location that minimizes the $H_2$ norm of the controlled system~\cite{natarajan2016actuator}. We find the optimal actuator location that minimizes the supremum of the cost of the controlled system described in Eq.~\eqref{eq:gl_cost} for all possible initial conditions. We employ brute force sampling of actuator locations while using the reduced-order model approximations.

Let the control input for the GL system (Eq.~\eqref{eq:sys_ginzlandau}) be given by the feedback law $u_k = -Kq_k$. The controlled dynamics and the cost function can be re-written as
\begin{align}\label{eq:cost_function}
    J =& \sum_{k=1}^{\infty} q_k^*(Q + K^*SK)q_k \\
    \text{s.t.} \quad & q_{k+1} = (A - BK)q_k \nonumber
\end{align}
for some given initial condition $q_0$. Let $\underline{Q} = (Q + K^*SK)$ and $\underline{A} = (A - BK)$. We are interested in the following infinite summation,
\begin{align*}
    J =& \;\sum_{k=1}^{\infty} q_0^*(\ul{A}^*)^k\ul{Q}\ul{A}^kq_0 \\
      =& \;q_0^*Fq_0.
\end{align*}
where $F = \sum_{k=1}^{\infty} (\ul{A}^*)^k\ul{Q}\ul{A}^k$. Notice that for a given dynamical system, the cost solely depends on the initial condition. When the controlled system governed by $\ul{A}$ is stable, i.e. all the eigenvalues of $\ul{A}$ lie within the unit circle, $F$ can be efficiently computed by solving the following discrete Lyapunov equation
\begin{align*}
    \ul{A}^*F\ul{A} - F + \ul{Q} = 0.
\end{align*}

The maximum value of the cost function is given by the largest eigenvalue of $F$ denoted by $\lambda_{\text{max}}(F)$. The initial condition that yields this cost is the eigenvector of $X$ that corresponds to the largest eigenvalue. We use the reduced-order approximation of the state transition matrix $A$ to construct low-order feedback gain $K$ and compute the values of $\lambda_{\text{max}}(F)$. We then conduct an exhaustive search over all possible actuator location in an interval to find the optimal actuator location. The performance of the reduced-order method is measured by how close this location is to the true optimal actuator location. 

\begin{figure}
\centering
\subfloat[]{\label{fig:actuator_bf_rank5}\includegraphics[width = 0.5\textwidth]{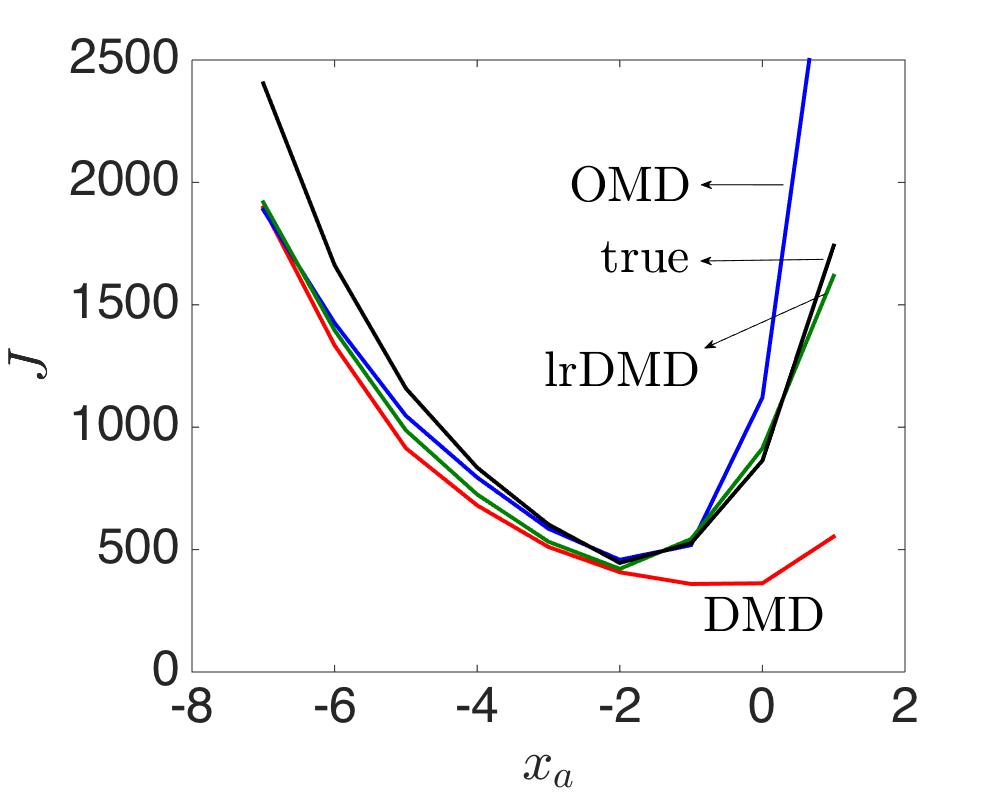}}
\subfloat[]{\label{fig:actuator_bf_rank9}\includegraphics[width = 0.5\textwidth]{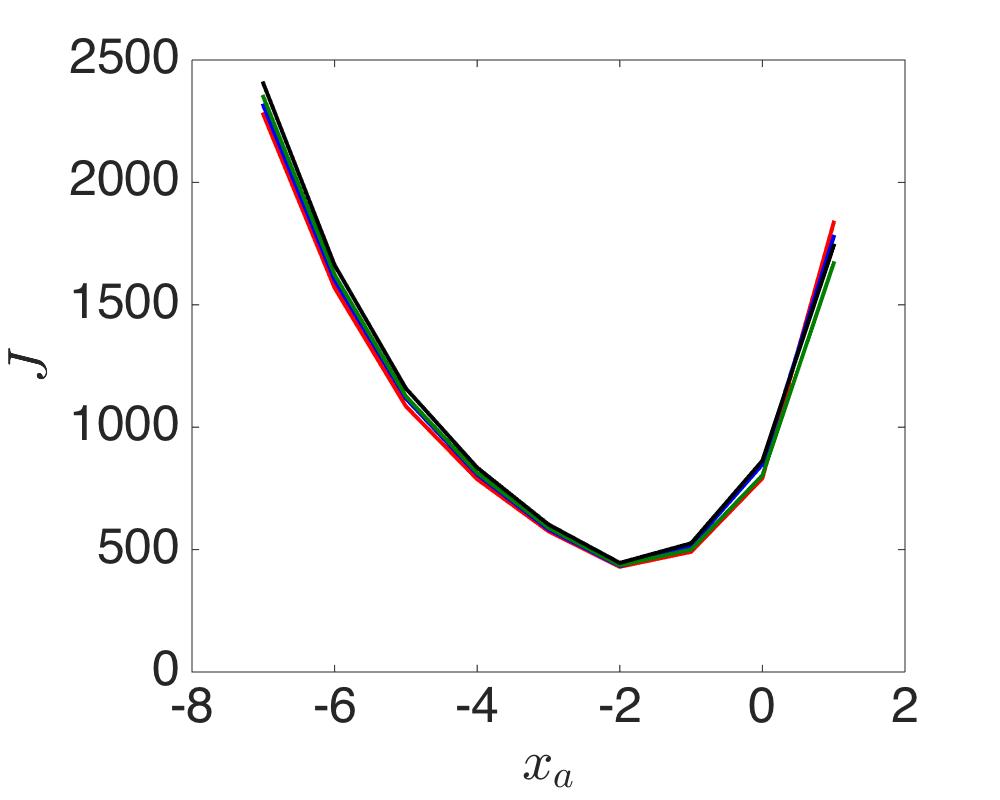}}
\caption{Supremum of the cost $J$ over all initial conditions for the controlled GL system with DMD (red), OMD (blue) and lrDMD (green) reduced-order controllers for (a) rank 5 approximation and (b) rank 9 approximation.}
\label{fig:actuator_loc}
\end{figure}

Figure~\ref{fig:actuator_loc} shows the supremum of the cost over all possible initial conditions for different actuator locations $x_a\in[-7,1]$ using DMD, OMD and lrDMD approximation compared with the true full-order system. The optimal actuator location is given by the minima of this plot. Figure~\ref{fig:actuator_bf_rank5} shows the results for a rank 5 approximation. lrDMD outperforms both DMD and OMD in matching the cost over the range of actuator locations considered. The DMD approximation attains its minima at $x_a = 0$ whereas the true optimal actuator location is $x_a = -2$. Optimal actuator locations given by OMD and lrDMD agree with the true optimal actuator location. Figure~\ref{fig:actuator_bf_rank9} shows the same results for a rank 9 approximation using DMD, OMD and lrDMD. At this rank all three methods perform equivalently and provide accurate costs for all actuator locations considered.
\subsection{Flow past a flat plate}\label{sec:ibpm}

In this section we demonstrate the performance of the controllers on stabilizing the two-dimensional uniform flow approaching an inclined flat plate. The freestream flow is at a low Reynolds number of $100$ and the flat plate is inclined at an angle of $35^{\circ}$. At these conditions, it has been shown~\cite{ahuja2010feedback} that the steady state of the flow is unstable and the flow exhibits periodic vortex shedding. The goal of the reduced-order controllers is to bring the system back to steady state from different initial conditions in the transition process.

The flow is simulated using the fast immersed boundary method developed in~\cite{colonius2008fast}. It is an efficient method to solve incompressible Navier-Stokes equations based on an immersed boundary formulation. In order to achieve uniform flow conditions in the far field, a multi-domain approach is employed. The domain of interest is considered to be embedded in a series of domains, each twice-as-large as the preceding but with the same number of the uniform grid points. The numerical parameters of the flow are taken to follow the work of~\cite{ahuja2010feedback}. The grid size used is $250\times 250$ and the domain of interest is given by $[-2,3] \times [-2.5,2.5]$ where the lengths are non-dimensionalized by the chord length of the flat plate, $L$. The center of the flat plate is located at the origin. Five domains, each with the same number of grid points are used for an effective computational domain that is $2^4$ times larger than the domain of interest. The time-step is taken as $dt = 0.01L/U_{\infty}$ where $U_{\infty}$ is the freestream velocity.

\subsubsection{Steady state}

\begin{figure} 
\centering
\subfloat[]{\label{fig:steady_fp}\includegraphics[height = 0.27\textwidth]{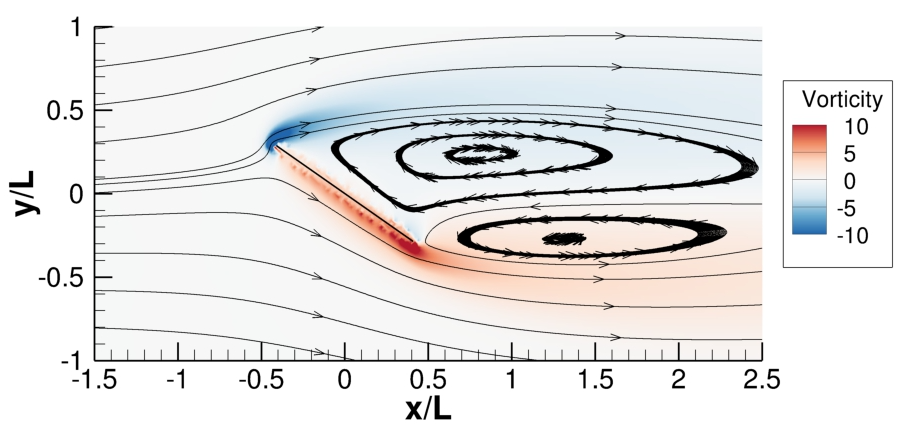}}
\subfloat[]{\label{fig:lcCL_fp}\includegraphics[height = 0.27\textwidth]{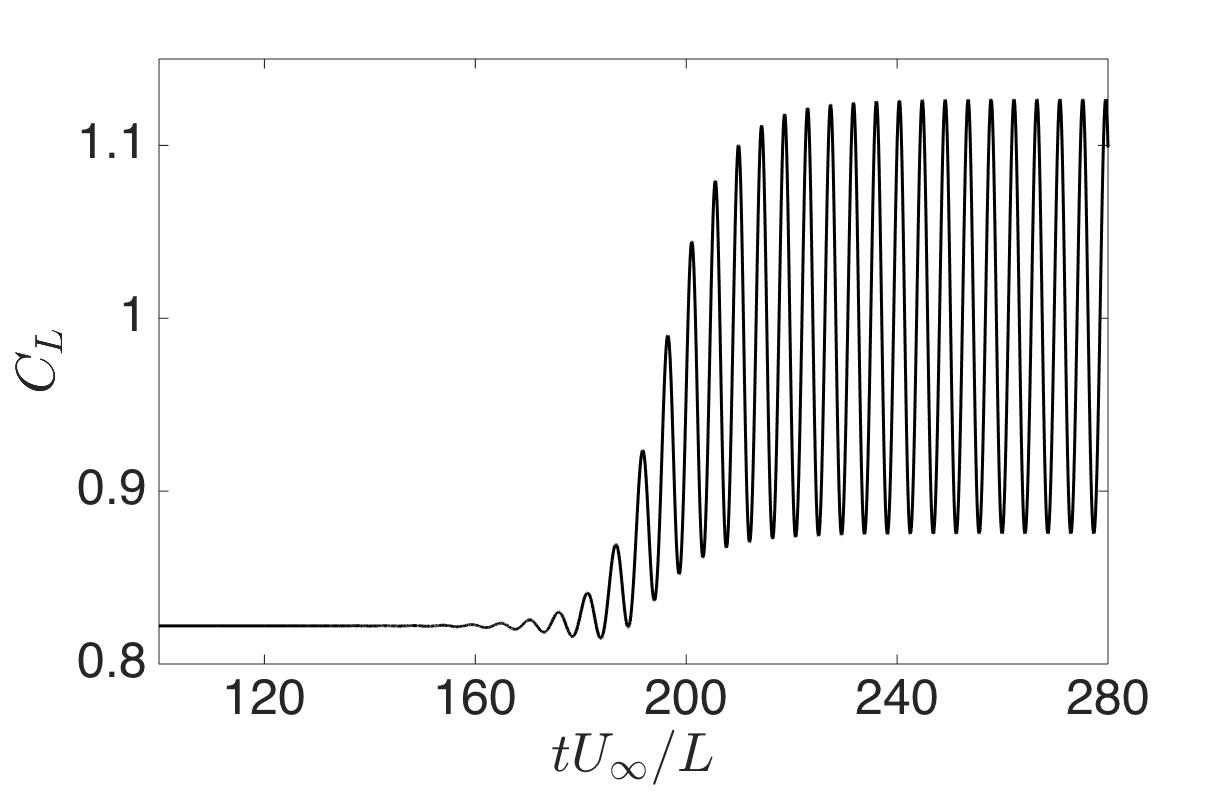}}
\caption{(a) Vorticity contours and velocity streamlines for steady flow over $35^{\circ}$ inclined flat plate (b) $C_L$ vs. time for $35^{\circ}$ inclined flat plate with the unstable steady state as the initial condition}
\label{fig:steadystate_fp}
\end{figure}

The computation of the steady state of an unstable flow is generally more difficult than a stable flow configuration. We are interested in the initial condition that is a fixed point of the governing equations of the flow~\cite{ahuja2009reduction}. Using a computational Fortran wrapper around the immersed boundary method code, we perform Newton-GMRES iterations on the nonlinear solver. We find the vorticity field $q$ that satisfies
\begin{align*}
    g(q) = q - \phi_T(q)
\end{align*}
where $\phi_T(\cdot)$ is the nonlinear solver that advances the solution $q_k$ at timestep $k$ to $q_{k+1} = \phi_T(q_{k})$ after $T$ timesteps. For the purpose of this study we choose $T = 50$ and iterate until a convergence tolerance of $10^{-8}$ in the $L_2$ norm. Figure~\ref{fig:steady_fp} shows the vorticity contours and velocity streamlines of the steady state while Figure~\ref{fig:lcCL_fp} shows the coefficient of lift as instabilities grow with the unstable steady state as the initial condition.

\subsubsection{Snapshots}

\begin{figure} 
\centering
\includegraphics[width = 0.8\textwidth]{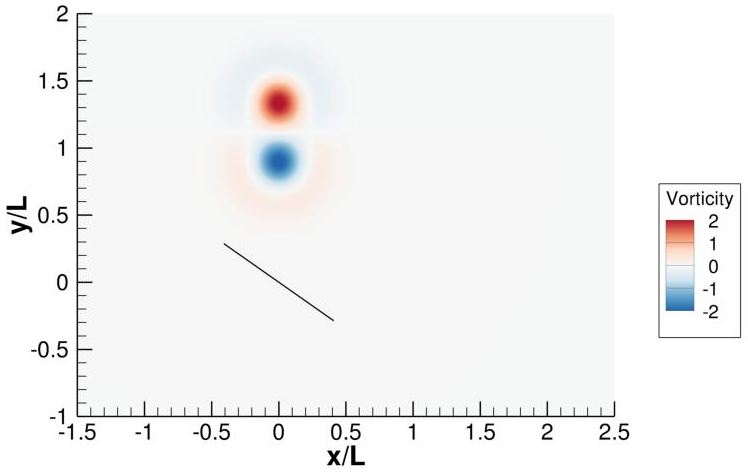}
\caption{Vorticity field generated by the actuator placed near the leading edge of a flat plate that is inclined at $35^{\circ}$ with the freestream.}
\label{fig:impulse_fp}
\end{figure}

The snapshots for the study are generated by the nonlinear impulse response of the flow to an actuator~\cite{ahuja2010feedback} located near the leading edge of the flat plate. The system can be described as
\begin{align}\label{eq:snapshots_eqn}
    q_{k+1} = \phi_T(q_k) + Bu_k
\end{align}
where $q_k$ is the vorticity field at iteration $k$, $\phi_T$ is the nonlinear solver that advances the vorticity field by $T$ timesteps, $B$ is the actuator and $u_k$ is the control input at iteration $k$. The actuator is a simple model of localized body force~\cite{ahuja2009reduction} at the actuator location near the leading edge of the flat plate. The instantaneous vorticity field generated by impulse control input of the actuator is
\begin{align*}
    B(r) = c[(1 - ar_1^2)\exp({-ar_1^2}) - (1 - ar_2^2)\exp({-ar_2^2})]
\end{align*}
where $r_i^2 = (x - x_{c,i})^2 + (y - y_{c,i})^2$ for $i={1,2}$. The constants $a$ and $c$ determine the shape and strength of the control, respectively. Table~\ref{tab:actuator_loc_fp} shows values of the constants used for this study. Figure~\ref{fig:impulse_fp} shows the vorticity field generated by the actuator relative to the position of the flat plate in the computational domain.

\begin{table} 
\begin{center}
\begin{tabular}{|c|c|c|c|c|c|}
\hline
 $x_{c,1}$ & $y_{c,1}$ & $x_{c,2}$ & $y_{c,2}$ & $a$ & $c$\\
 \hline
 0 & 1.3423 & 0 & 0.89 & 20 & 2\\
 \hline
\end{tabular}
\end{center}
\caption{Numerical parameters for the actuator location and strength for feedback control of flow past inclined flat plate.}
\label{tab:actuator_loc_fp}
\end{table}

\begin{figure} 
\centering
\includegraphics[width =0.8\textwidth]{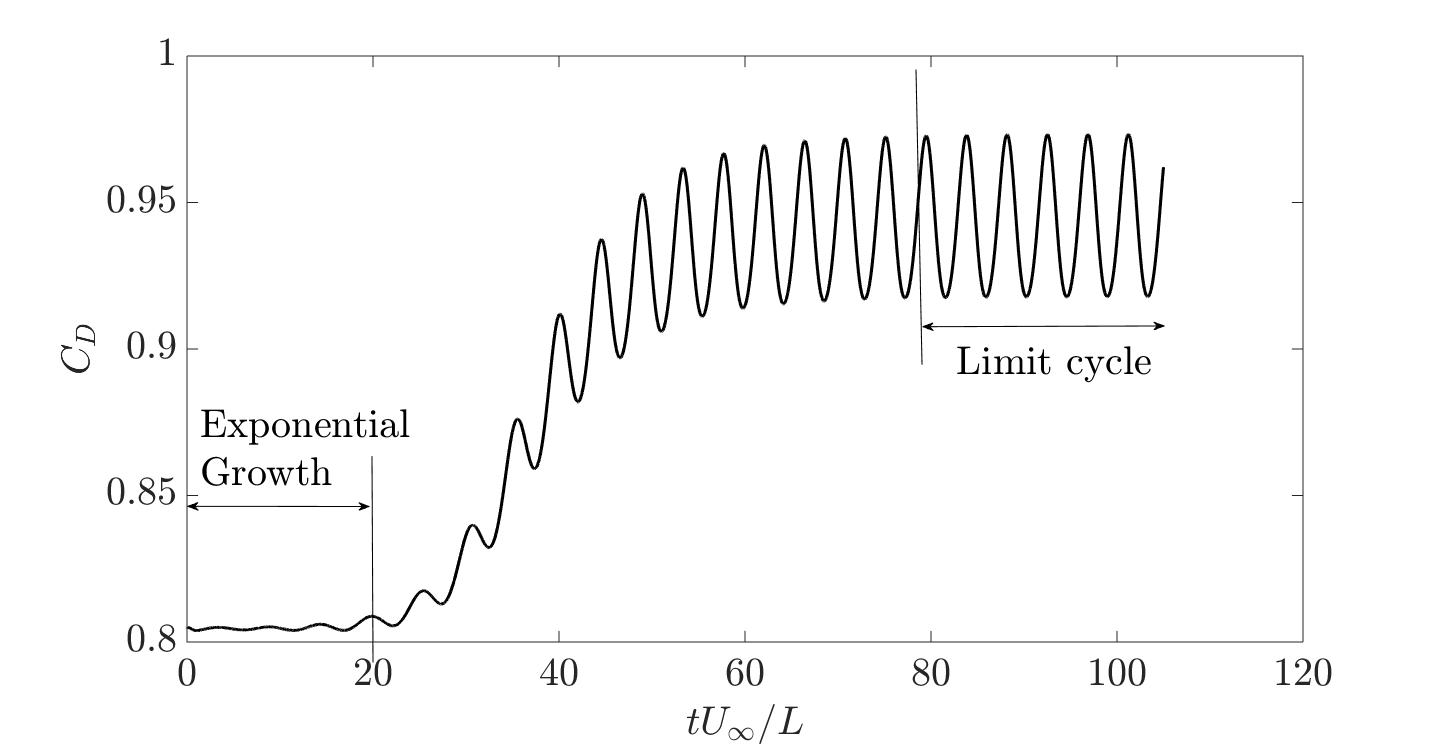}
\caption{$C_D$ vs. time plot of the impulse response of $35^{\circ}$ inclined flat plate with steady state as the initial condition. Snapshots of this simulation were used in the study.}
\label{fig:impulse_cl_fp}
\end{figure}

Getting a set of snapshots that captures the essential physics is crucial for the performance of data-driven methods. Therefore, an important step in building data-driven reduced-order controllers is to identify the nature of instabilities in the flow and select data that best captures the behaviour of interest. The goal is to select a set of snapshots that capture the near-linear behaviour of the flow before the nonlinearities dominate the system dynamics. Figure~\ref{fig:impulse_cl_fp} shows the evolution with time of drag coefficient of the impulse response of the flat plate. The two regions marked as exponential growth and limit cycle can be approximated by linear dynamics. We are only interested in the exponential growth region and we choose to sample this region for our study.
\subsubsection{Closed loop control}

We model the system as linear about the steady state of the flow. To this end, we subtract the computed steady state from the flow snapshots and build a reduced-order model for the perturbations about the steady state. The reduced-order model can be described as
\begin{align*}
    \tilde{q}_{k+1} = A\tilde{q}_{k} + Bu_{k}
\end{align*}
where $\tilde{q}_k = q_{k} - \overline{q}$ and $\overline{q}$ is the steady state vorticity field. This can be considered as a model of the dynamics of perturbations in the flow around the steady state.

The control is performed using a Fortran wrapper around the nonlinear solver. The controller is activated at three different points in the transition process, $t_0 = \{170, 190, 210\}$ (refer to Fig.~\ref{fig:lcCL_fp}), in separate simulations, to show the effect of the nonlinear dynamics of the flow. The nonlinear effects grow larger as the flow deviates from the steady state with time. At $t_0 = 170$, the flow shows exponential growth in perturbation which is predominantly linear behavior. At $t_0 = 190$, the flow has entered an algebraic growth in perturbation magnitude and $t_0 = 210$ is just before the flow enters a limit cycle behavior. At $t = \{190, 210\}$ the perturbations are too large for the linear assumption to be valid.

\begin{figure} 
\centering
\subfloat[]{\label{fig:170_20dt_control}\includegraphics[width = 0.5\textwidth]{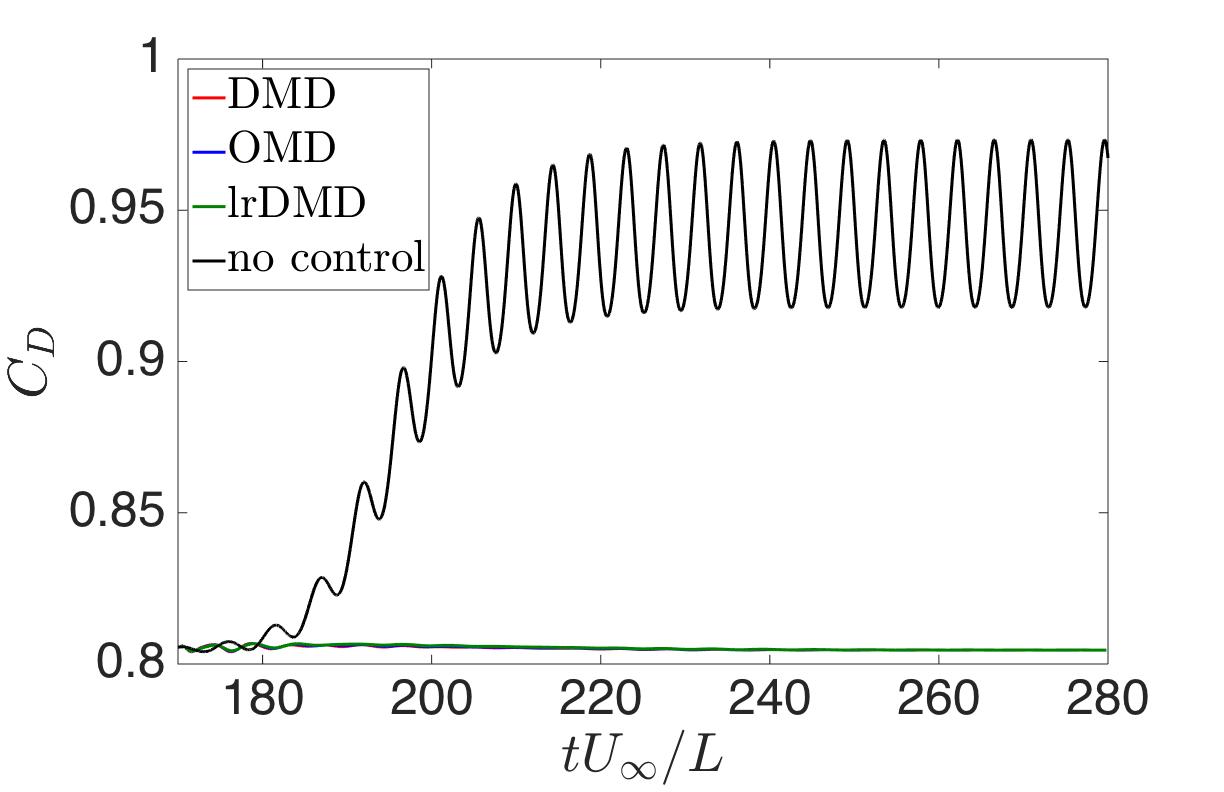}}\\
\subfloat[]{\label{fig:190_20dt_control}\includegraphics[width = 0.5\textwidth]{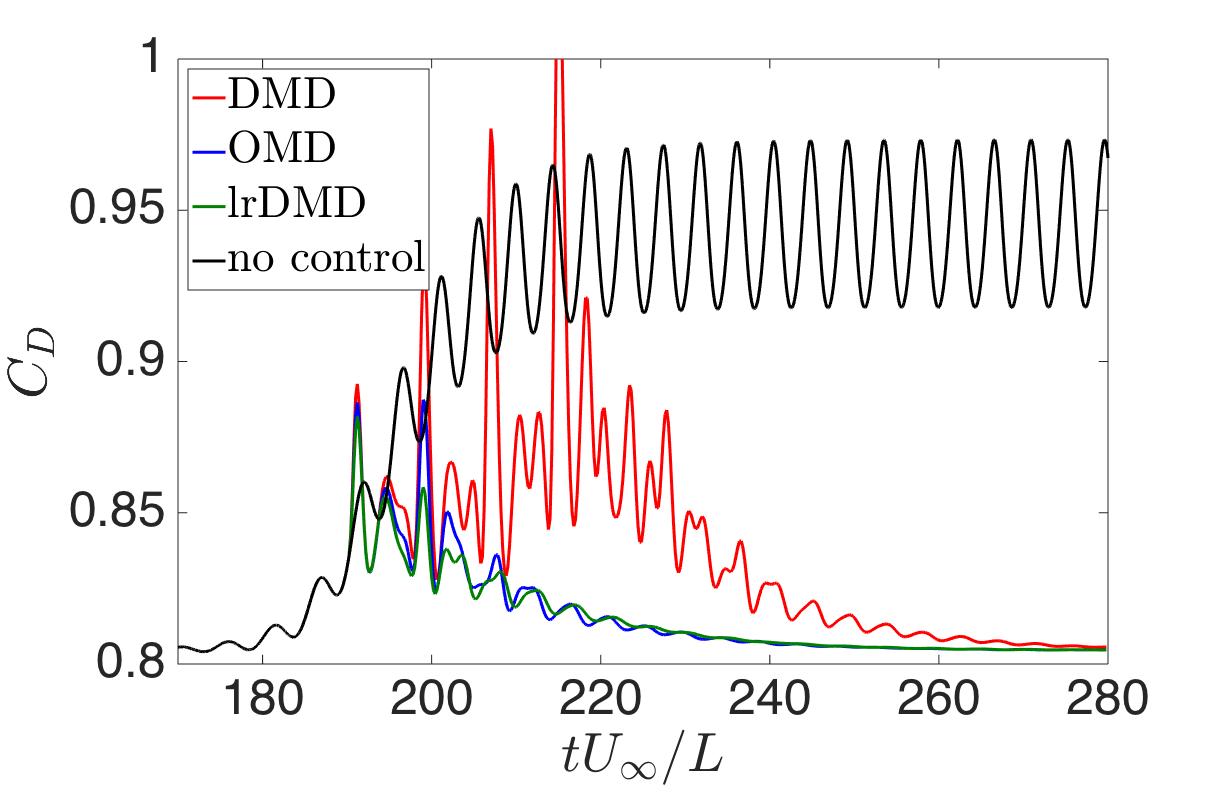}}
\subfloat[]{\label{fig:210_20dt_control}\includegraphics[width = 0.5\textwidth]{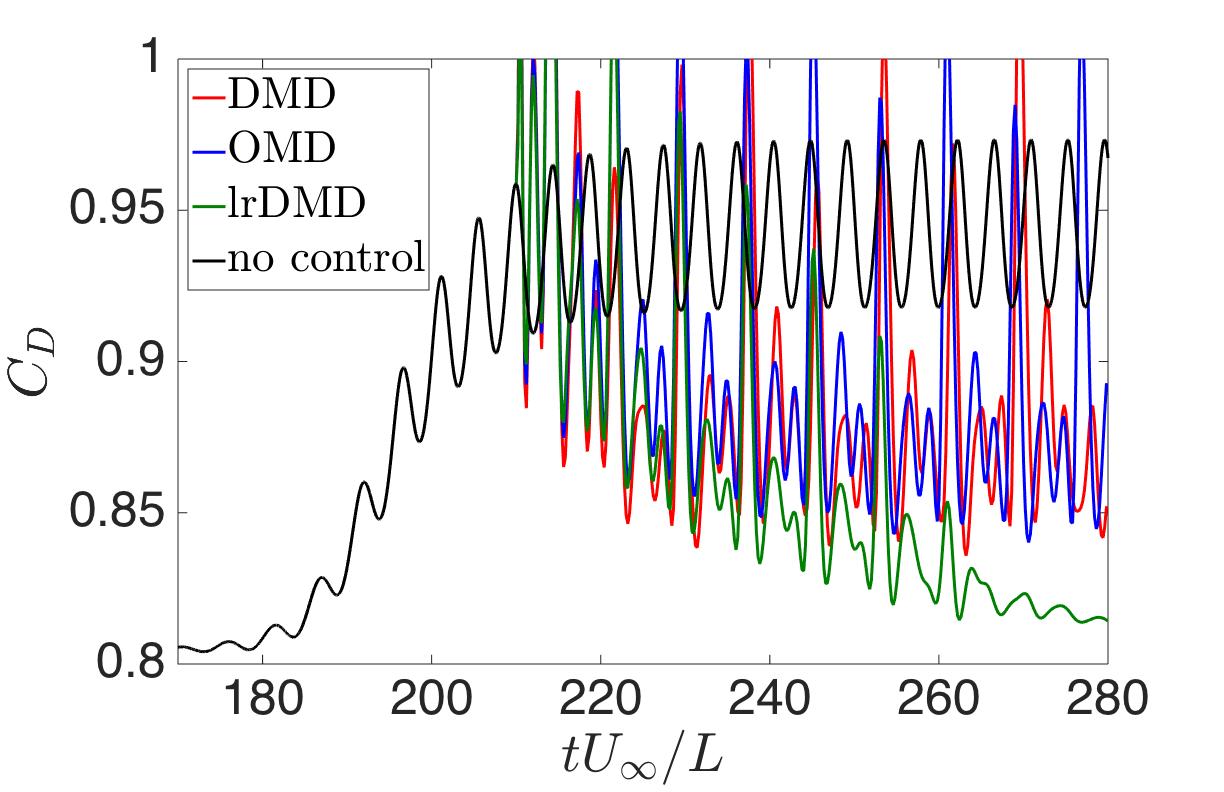}}
\caption{$C_D$ vs. time for $35^{\circ}$ inclined flat plate with the actuator activated at (a) $t_0 = 170$ (b) $t_0 = 190$ and (c) $t_0 = 210$ using DMD, OMD and lrDMD based reduced-order controllers.}
\label{fig:20dt_control}
\end{figure}

Equation~\eqref{eq:snapshots_eqn} is used to generate snapshots to build the data matrix. We choose $T = 20$ and collect $200$ snapshots to cover multiple period of oscillations of the exponentially growing perturbations. Linear approximations of rank $20$ using DMD, OMD and lrDMD are used to generate the LQR controllers. Since $T = 20$, the feedback control is applied after every $20$ timesteps of the nonlinear flow solver. A rank of $20$ is high enough to faithfully reconstruct the dynamics but much lower than the dimension of the full order system which is of $O(10^4)$. Figure~\ref{fig:20dt_control} shows the evolution of $C_D$ as the controller is switched on at the three time points. Figure~\ref{fig:170_20dt_control} shows DMD, OMD and lrDMD are all able to prevent the growth of perturbations in the flow when the controller is activated at $t_0 = 170$. This is expected since the perturbations are small and the flow is in the linear regime. In Fig.~\ref{fig:190_20dt_control} at $t_0 = 190$, OMD and lrDMD show very similar control performance and are able to suppress the vortex shedding earlier than DMD. Finally in Fig.~\ref{fig:210_20dt_control}, lrDMD clearly outperforms both OMD and DMD in bringing the flow back to the steady state. This shows that in the linear regime all three methods are equivalent in performance. As the perturbations grow and nonlinear effects in the flow become significant, the additional flexibility of lrDMD that allows a different input and output subspaces results in a better control performance compared to OMD and DMD.

\begin{figure}
\centering
\subfloat[]{\label{fig:170_20dt_strength}\includegraphics[width = 0.5\textwidth]{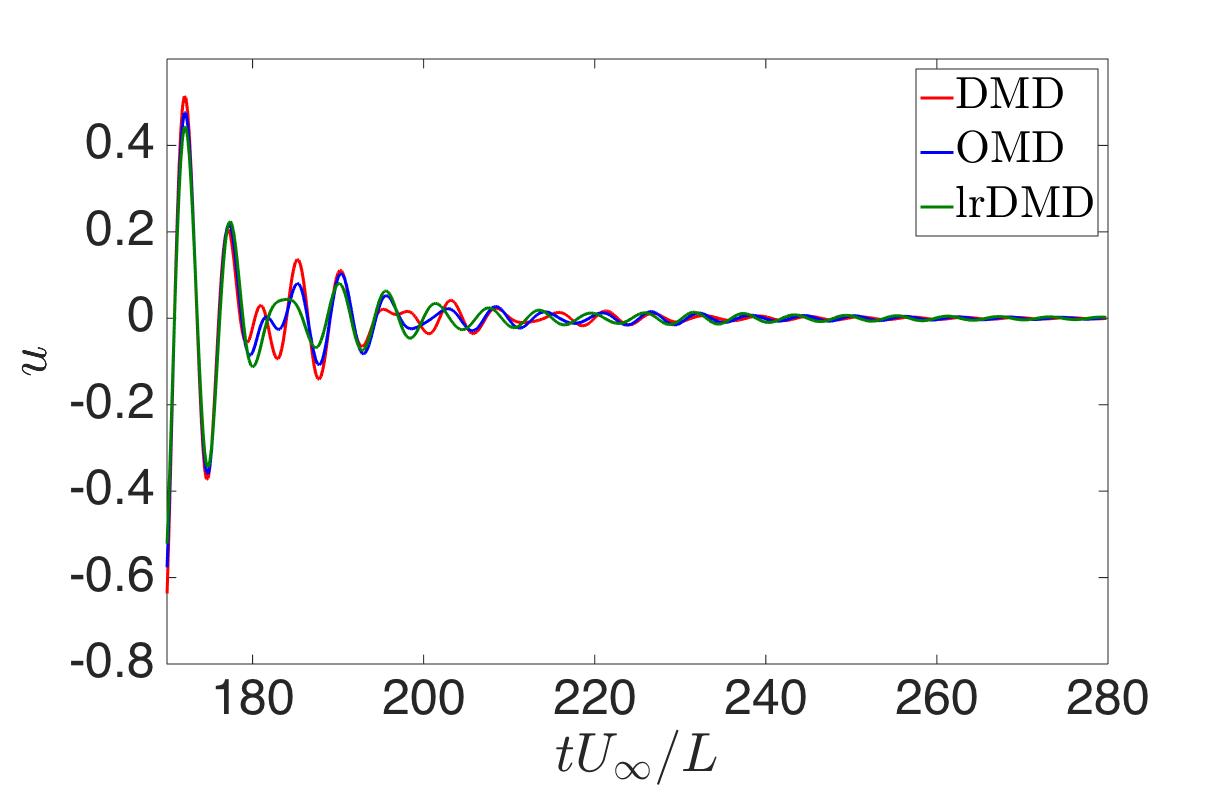}}\\
\subfloat[]{\label{fig:190_20dt_strength}\includegraphics[width = 0.5\textwidth]{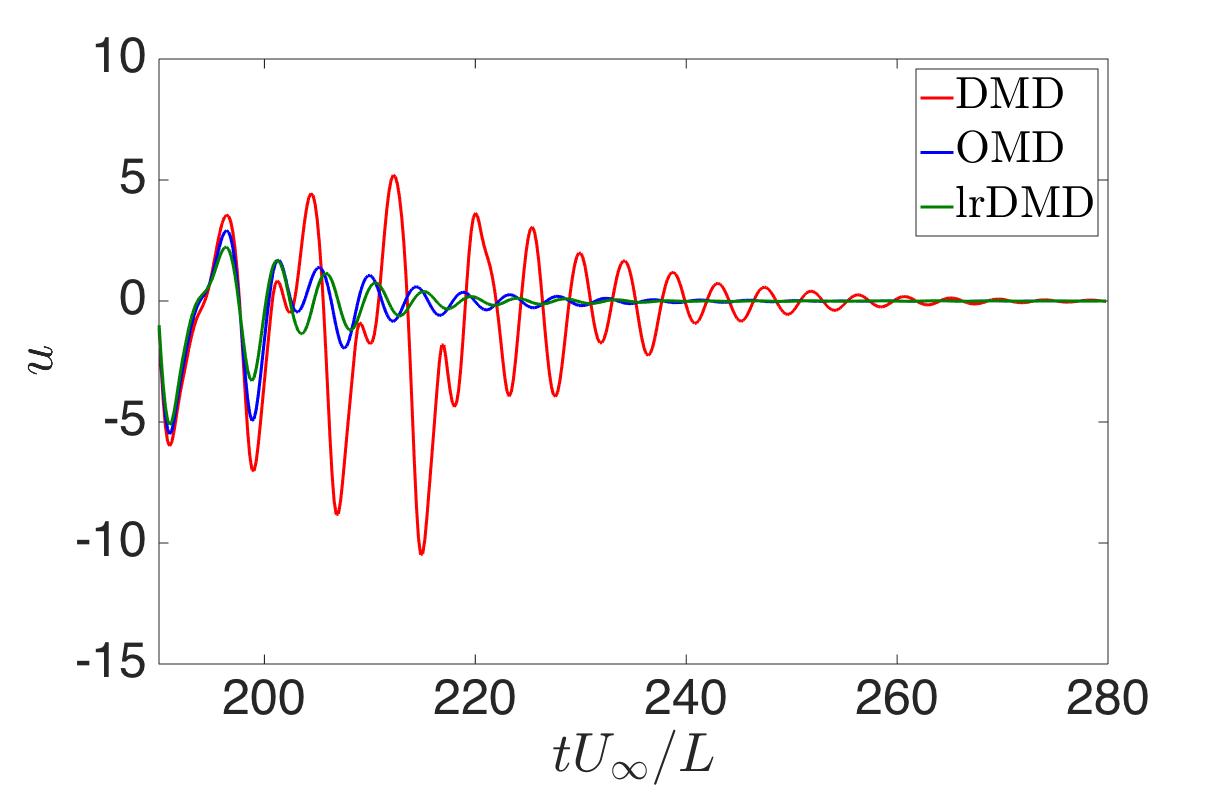}}
\subfloat[]{\label{fig:210_20dt_strength}\includegraphics[width = 0.5\textwidth]{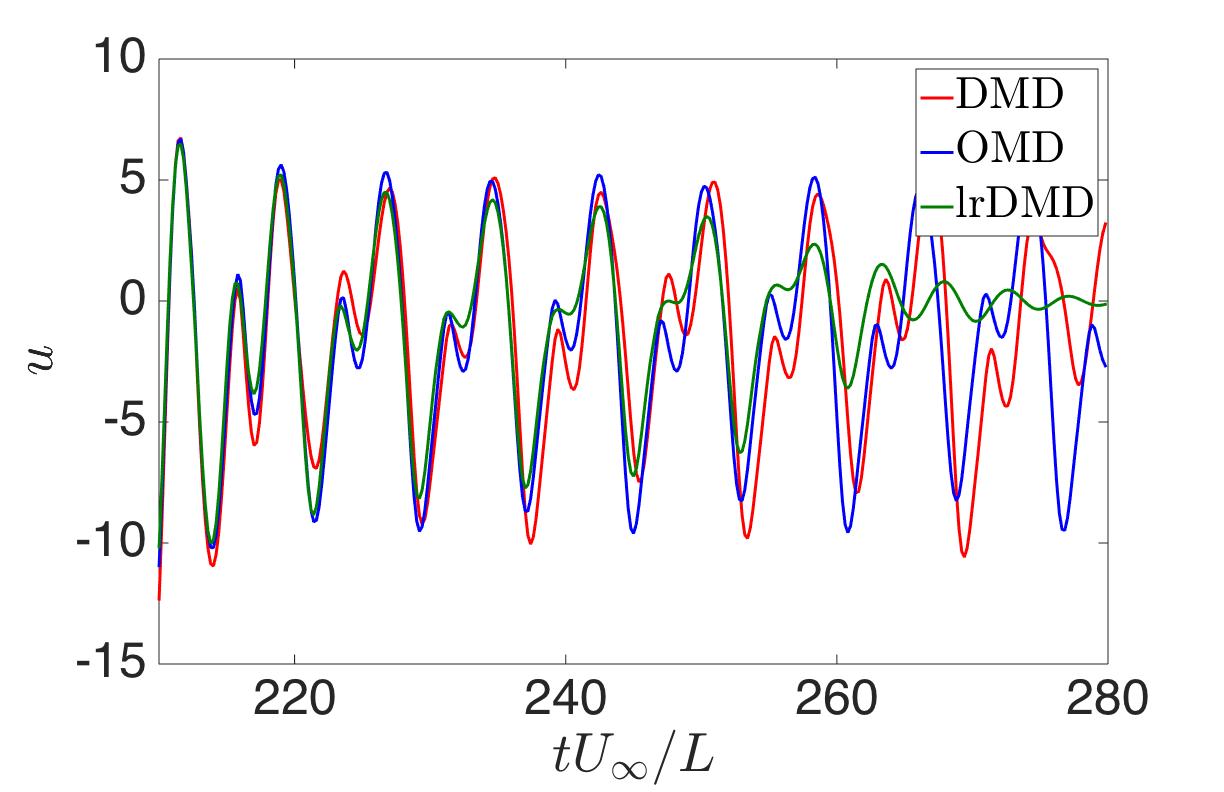}}
\caption{Strength of the actuation when the controller was switched on at (a) $t_0 = 170$ (b) $t_0 = 190$ and (c) $t_0 = 210$ for a $35^{\circ}$ inclined flat plate using DMD, OMD and lrDMD based reduced-order controllers.}
\label{fig:20dt_strength}
\end{figure}

Figure~\ref{fig:20dt_strength} shows the actuation of the controller with time for the three activation times of the controllers using DMD, OMD and lrDMD used in Fig.~\ref{fig:20dt_control}. The results follow the same trend as Fig.~\ref{fig:20dt_control}. In Fig.~\ref{fig:170_20dt_strength} when the controller is activated at $t_0 = 170$, all three controllers show similar performance. At $t_0 = 190$ in Fig.~\ref{fig:190_20dt_strength}, OMD and lrDMD incur much lower cost compared to DMD and in Fig.~\ref{fig:210_20dt_strength} the cost with lrDMD controller is clearly much lower than DMD and OMD controllers.
\begin{table}
    \centering
        \begin{tabular}{|c|c|c|c|c|c|c|}
            \hline
             \multirow{2}{4em}{Method} & \multicolumn{2}{|c|}{$t_0 = 170$} & \multicolumn{2}{|c|}{$t_0 = 190$} & \multicolumn{2}{|c|}{$t_0 = 210$} \\
             \cline{2-7}
                 &  $T=50$ & $T=100$ &  $T=50$ & $T=100$ & $T=50$ & $T=100$ \\
             \hline
             DMD & \cmark & \cmark & \cmark & \xmark & \xmark & \xmark \\
             \hline
             OMD & \cmark & \cmark & \cmark & \cmark & \cmark & \xmark \\
             \hline
           lrDMD & \cmark & \cmark & \cmark & \cmark & \cmark & \xmark\\
             \hline
        \end{tabular}
    \caption{Summary of the feedback control performance of DMD, OMD and lrDMD based controllers for flow over inclined flat plate. The time at which the controller was activated is indicated by $t_0$ and $T = \{50, 100\}$ refers to the time separation of corresponding data snapshots in the data matrix used to construct the controllers. \cmark$\;$ indicates that the flow converged to steady state and \xmark$\;$ indicates that the flow became unstable.}
    \label{tab:checklist}
\end{table}
\ignore{
\begin{table}
    \centering
        \begin{tabular}{|c|c|c|c|c|c|c|c|c|c|}
            \hline
             \multirow{2}{4em}{Method} & \multicolumn{3}{|c|}{$t_0 = 170$} & \multicolumn{3}{|c|}{$t_0 = 190$} & \multicolumn{3}{|c|}{$t_0 = 210$} \\
             \cline{2-10}
                 &  $T=20$ &  $T=50$ & $T=100$ &  $20$ &  $50$ & $100$ &  $20$ &  $50$ & $100$ \\
             \hline
             DMD & \cmark & \cmark & \cmark & \cmark & \cmark & \xmark & \cmark & \xmark & \xmark \\
             \hline
             OMD & \cmark & \cmark & \cmark & \cmark & \cmark & \cmark & \cmark & \cmark & \xmark \\
             \hline
           lrDMD & \cmark & \cmark & \cmark & \cmark & \cmark & \cmark & \cmark & \cmark & \xmark\\
             \hline
        \end{tabular}
    \caption{Summary of the feedback control performance of DMD, OMD and lrDMD based controllers for flow over inclined flat plate. $t_0$ indicates the time at which the controller was activated and $T = \{20, 50, 100\}$ refers to the time separation of corresponding data snapshots in the data matrix used to construct the controllers. \cmark$\;$ indicates that the flow converged to steady state and \xmark$\;$ indicates that that the flow became unstable.}
    \label{tab:checklist}
\end{table}
}
\par
We repeat this numerical experiment using data matrices with consecutive snapshots that are $50$ and $100$ iterations apart. Table~\ref{tab:checklist} shows the results of feedback based control at different initial conditions of the flow for DMD, OMD and lrDMD based controllers. As the separation between the flow snapshots increases, the actuation is being applied further apart in time which makes it harder to suppress the vortex shedding process. OMD and lrDMD controllers are successful in controlling the flow at all conditions except when the snapshots are $100$ timesteps apart and the initial condition is at $t_0 = 210$ at which the flow becomes unstable. DMD however fails more often especially when the snapshots are $100$ timesteps apart, it is only able to control when $t_0 = 170$.
\section{Conclusion}\label{sec:conclusion}

To summarize, we introduce a method to approximate the dynamics of an unsteady fluid flow by a rank-constrained linear representation. We solve for the optimal linear map of a fixed user defined rank that reconstructs the data snapshots by minimizing the $L_2$ norm of the reconstruction error. Two methods of solving the optimization problem are presented and their limitations are discussed. The first method is a subspace projection method that is very fast and provides a good approximation to the optimal solution of the optimization problem. The second method employs a gradient descent approach which is guaranteed to converge at a local minimum. Reduced-order models generated using lrDMD are shown to produce lower reconstruction errors compared to Optimal Mode Decomposition (OMD) and Dynamic Mode Decomposition (DMD) while incurring comparable computational times. These reduced-order models are used to construct low-rank full-state feedback controllers to control model fluid flows. 

We employ LQR based feedback control using the reduced-order models on the linearized Ginzburg-Landau equation in the globally unstable regime. lrDMD is able to stabilize the system at a much lower rank approximation as compared to DMD. The reduced-order models are also used to find optimal actuator location using brute force sampling approach. The true optimal actuator location is obtained using lrDMD at a rank 5 approximation, whereas DMD requires a rank 9 approximation to provide the true optimal actuator location. We also employed LQR based feedback control on unsteady flow over an inclined flat plate. OMD and lrDMD perform equally well in controlling the flow whereas DMD incurs much higher costs in the control for all cases.

This paper shows the potential of using a linear map that is a mapping between different subspaces to construct reduced-order controllers for unsteady high-dimensional systems. We observe that these linear maps fit the data better and can lead to better performing reduced-order controllers.

\section*{Acknowledgements}
The authors would like to thank Mr. Daniel Floryan and Dr. Clarence Rowley for their help with the code used to simulate flow past an inclined flat plate. This work was sponsored, in part, by the Office of Naval Research (ONR) as part of the Multidisciplinary University Research Initiatives (MURI) Program, under grant number N00014-16-1-2617.

\appendix

\section{Numerical methods}\label{app:methods}

In this section we present the two methods to solve the lrDMD optimization problem.

\subsection{Subspace projection method}\label{app:sub_proj}

In this method, we use iterative subspace projection to find a good approximation of the optimal solution. We first make the observation that $Q_R$ is an orthogonal projection matrix in the column space of $X^TR$. This means that there exists an orthogonal matrix $C_R$ such that $C_RC_R^T = Q_R$ and $\text{Im}(C_R) = \text{Im}(Q_R)$. Substituting $Q_R$ with $C_RC_R^T$, we get the following cost function,
\begin{align*}
    G(L,R) = -\norm{L^TYC_R}_F^2.
\end{align*}
with the constraint that $\text{Im}(C_R) = \text{Im}(Q_R)$. If $C_R$ is fixed, the optimal solution for $L$ is given by the left singular vectors of $YC_R$. Finding $C_R$ for a fixed $L$ under the given constraint is not as trivial. If the left singular vectors of $X^TR$ span the same space as the left singular vectors of $Y^TL$, then optimal $C_R$ under the constraint will just be the left singular vectors of $X^TR$. As an approximation, we try to find $R$ that minimizes the distance between $X^TR$ and $Y^TL$ by solving the following `Orthogonal Procrustes Problem'~\cite{schonemann1966generalized}

\begin{align*}
    \min_R \norm{Y^TL - X^TR}_F^2.
\end{align*}

The closed form optimal solution to this problem is $R = UV^T$ where $U$ and $V$ are the left and the right singular vector matrices respectively from the singular vector decomposition of $XY^TL$. $C_R$ is given by the orthogonal basis of the column space of $X^TR$ denoted by $\Pi(X^TR)$. Algorithm~\ref{alg:subspace} describes all the steps for this method.

The solution provided by this algorithm relies heavily on the initial guess. We only need an initial guess for $R$ or, effectively, $C_R$. In our study, we choose the $r$ leading left singular vectors of $X$ as the initial guess for $R$. Note that this initial guess is the same as the projection subspace used in DMD. Due to the optimal choice of $L$ for a fixed $C_R$, the chosen initial guess ensures that the algorithm provides a solution with a reconstruction error at most as high as the error in DMD reconstruction of the same rank.

\begin{algorithm}
\caption{Subspace projection method}
\label{alg:subspace}
\begin{algorithmic}[1]
\REQUIRE $Y\in \mb{R}^{m\times n}$, $X\in \mb{R}^{m\times n}$
\STATE Guess initial $L_0$,$R_0$ and compute $C_{R_0} \leftarrow \Pi(X^TR_0)$
\STATE $k = 0$
\REPEAT
\STATE $L_{k+1} \leftarrow \Pi(YC_{R_{k}})$
\STATE $R_{k+1} \leftarrow \argmin\norm{Y^TL_{k}-X^TR}_F^2$
\STATE $C_{R_{k+1}} \leftarrow \Pi(X^TR_{k+1})$
\STATE $\epsilon \leftarrow (G(L_{k+1},R_{k+1})-G(L_{k},R_{k}))/G(L_{k},R_{k})$
\STATE $k \leftarrow k + 1$
\UNTIL $\epsilon \leq \text{threshold}$
\STATE $D = (L_{k}^TYX^TR_{k})(R_{k}^TXX^TR_{k})^{-1}(R_{k}^TL_{k})$
\RETURN $L_{k},D,R_{k}$
\end{algorithmic}
\end{algorithm}

\subsection{Gradient descent method}\label{app:gradD}

In this section, we describe the Riemannian gradient descent method employed to solve the optimization problem. For a thorough review on Riemannian optimization on the Grassmanian manifold the reader is referred to~\cite{edelman1998geometry}. For a function $F(L)$ defined on the Grassmanian manifold where $L\in \mc{G}_{r,m}$, the gradient $\text{grad}\, F(L) \in \mc{T}_{L}\mc{G}_{r,m}$ (tangent space to the manifold at the point $L$) is defined as
\begin{align}\label{eq:grad}
    \text{grad}\,F(L) &= \nabla F(L) - LL^T\nabla F(L) \\
    \text{where}\quad (\nabla F(L))_{ij} &= \frac{d F}{dL_{ij}}
\end{align}

Similarly, the action of the Hessian on any tangent vector $dL \in \mc{T}_L\mc{G}_{r,m}$ is defined as
\begin{align}\label{eq:hess}
    \text{Hess}\,F(L)[dL] &= (I_m - LL^T)\nabla^2F(L)[dL] - dL(L^T\nabla F(L)) \\
    \text{where}\quad (\nabla^2 F(L))_{ij,kl} &= \frac{d^2F}{dL_{ij}dL_{kl}}.
\end{align}
 
Further details of the method and its implementation can be seen in~\cite{sashittal2018low}. We employ the Trust-region algorithm~\cite{absil2007trust} to solve the optimization problem using the MATLAB package ManOpt~\cite{boumal2014manopt}.

\bibliographystyle{unsrt}
\bibliography{reference.bib}

\end{document}